# Detecting and Constraining $N_2$ Abundances in Planetary Atmospheres Using Collisional Pairs

Edward W. Schwieterman[1,2,3], Tyler D. Robinson[2,4], Victoria S. Meadows[1,2,3], Amit Misra[1,2,3], and Shawn Domagal-Goldman[2,5]

[1]Astronomy Department, University of Washington, Seattle, WA 98115, USA; eschwiet@uw.edu

[2]NAI Virtual Planetary Laboratory, Seattle, WA 98115, USA

[3]Astrobiology Program, University of Washington, Seattle, Washington, USA

[4]NASA Ames Research Center, Moffet Field, CA 94035, USA

[5]NASA Goddard Space Flight Center, Greenbelt, MD 20771, USA

ABSTRACT

Characterizing the bulk atmosphere of a terrestrial planet is important for determining surface pressure and potential habitability. Molecular nitrogen ($N_2$) constitutes the largest fraction of Earth's atmosphere and is likely to be a major constituent of many terrestrial exoplanet atmospheres. Due to its lack of significant absorption features, $N_2$ is extremely difficult to remotely detect. However, $N_2$ produces an $N_2$-$N_2$ collisional pair, $(N_2)_2$, which is spectrally active. Here we report the detection of $(N_2)_2$ in Earth's disk-integrated spectrum. By comparing spectra from NASA's *EPOXI* mission to synthetic spectra from the NASA Astrobiology Institute's Virtual Planetary Laboratory three-dimensional spectral Earth model, we find that $(N_2)_2$ absorption produces a ~35% decrease in flux at 4.15 μm. Quantifying $N_2$ could provide a means of determining bulk atmospheric composition for terrestrial exoplanets and could rule out abiotic $O_2$ generation, which is possible in rarefied atmospheres. To explore the potential effects of $(N_2)_2$ in exoplanet spectra, we used radiative transfer models to generate synthetic emission and transit transmission spectra of self-consistent $N_2$-$CO_2$-$H_2O$ atmospheres, and analytic $N_2$-$H_2$ and $N_2$-$H_2$-$CO_2$ atmospheres. We show that $(N_2)_2$ absorption in the wings of the 4.3 μm $CO_2$ band is strongly dependent on $N_2$ partial pressures above 0.5 bar and can significantly widen this band in thick $N_2$ atmospheres. The $(N_2)_2$ transit transmission signal is up to 10 ppm for an Earth-size planet with an $N_2$-dominated atmosphere orbiting within the HZ of an M5V star and could be substantially larger for planets with significant $H_2$ mixing ratios.

# 1. INTRODUCTION

Molecular nitrogen ($N_2$) comprises the bulk of Earth's modern atmosphere, is an important constituent in the atmospheres of Venus and Mars, and dominates the atmosphere of Saturn's icy moon Titan. Because of its geochemical stability, $N_2$ has constituted the largest fraction of Earth's atmosphere throughout its history, though its total abundance may have varied over time (Goldblatt et al. 2009; Som et al. 2012; Marty et al. 2013; Johnson & Goldblatt 2015). While $N_2$ is the most abundant gas in Earth's atmosphere, it is extremely difficult to detect remotely, because it is a symmetric homonuclear molecule with no transitional dipole moment and therefore does not produce spectral features throughout the visible and near-infrared portions of the spectrum. $N_2$ has a significant dissociation cross section in the Extreme Ultraviolet (EUV), $\lambda < 0.1$ μm (Samson et al. 1987; Stark et al. 1992), but the planetary flux at these wavelengths is likely to be low both due to a lack of EUV light from the incident stellar spectrum and significant absorption in the planet's atmosphere. Similarly, $N_2$-$N_2$ collisionally-induced absorption (CIA) in the far infrared ($\lambda > 40$ μm) (Borysow & Frommhold 1986) will be difficult to discern from many planetary spectra because it can be masked by water vapor absorption. However, $N_2$-$N_2$ collisions create short-lived $N_2$-$N_2$ pairs, also written as $(N_2)_2$, which are spectrally active at shorter wavelengths. $(N_2)_2$ has an absorption feature near 4.3 μm, nearly coincident with the 4.3 μm $CO_2$ band, but with broader wings at Earth-like $N_2$ abundances. The $(N_2)_2$ absorption in this wavelength region is dominated by CIA, but $N_2$ quadrupole line absorption, and absorption by true $(N_2)_2$ van der Waals molecules also contributes (Long et al. 1973). The temperature and wavelength-dependent absorption coefficients for this primarily collisional feature have been measured empirically (Farmer & Houghton 1966; Lafferty et al. 1996), and $(N_2)_2$ absorption has been observed in solar occultation (transmission) observations by Earth-observing satellites (Rinsland 2004), but, to our knowledge, it has never before been considered in the context of planet characterization from full-disk or transit transmission observations.

A comprehensive study of a planetary atmosphere would require determination of its bulk properties, such as atmospheric mass and composition, which are crucial for ascertaining surface conditions. Because $(N_2)_2$ is detectable remotely, it can provide an extra tool for terrestrial planet characterization. For example, the level of $(N_2)_2$ absorption could be used as a pressure metric if $N_2$ is the bulk gas, and break degeneracies between the abundance of traces gases and the foreign pressure broadening induced by the bulk atmosphere. If limits can be set on surface pressure, then the surface stability of water may be established if information about surface temperature is available. The technique of using collisional pairs (or dimers) to determine atmospheric pressure in the reflection and transmission spectra of exoplanets was first described by Misra et al. (2014a), who considered $(O_2)_2$ features specifically. The absorption by CIA or dimer molecules is quadratic in number density, while it is linear for monomers, making CIA or dimers especially sensitive to different maximum gas densities. Misra et al. (2014a) considered the strength of $(O_2)_2$ absorption features at 1.06 μm and 1.27 μm specifically as pressure discriminators. However, for almost half (Kump 2008), or possibly more (Planavsky et al. 2014), of Earth's geologic history, there was little $O_2$ present in the atmosphere, as oxygenic photosynthesis had not yet evolved or geochemical sinks for $O_2$ had not yet been exhausted (Buick 2008). For a substantial fraction of this time period life existed on the Earth (Brasier et al. 2015; Wacey et al. 2011). Therefore, if Earth is taken as an informative example, there may be many habitable and inhabited planets with no detectable $O_2$. For these worlds pressure determination with $(O_2)_2$ will be impossible and some other metric must be employed, such as quantifying $(N_2)_2$.

Additionally, it has been shown that atmospheres with low abundances of non-condensable gases such as $N_2$ and argon (Ar) are more susceptible to the abiotic accumulation of $O_2$, creating a potential false positive for life (Wordsworth & Pierrehumbert 2014). This occurs when the surface temperature is high enough, and the non-condensable gas abundance low enough, for water vapor to dominate the lower atmosphere. A water dominated atmosphere lacks a cold trap, allowing water to more easily diffuse into the stratosphere and become photodissociated, leaving free $O_2$ to build up over time. Direct detection of $N_2$ (through $(N_2)_2$) could rule out abiotic $O_2$ via this mechanism and, in tandem with detection of significant $O_2$ or $O_3$, potentially provide a robust biosignature. Moreover, the simultaneous detection of $N_2$, $O_2$, and a surface ocean would establish the presence of a significant thermodynamic chemical disequilibrium (Krissansen-Totton et al. 2015) and further constrain the false positive potential. Finally, $(N_2)_2$ overlaps significantly with the 4.3 μm $CO_2$ band, which may be used in future exoplanet observations to quantify $CO_2$ abundances. This retrieval would be biased unless the absorption in the wings from $(N_2)_2$ is considered. This effect is accounted for by Earth-observers using solar occultation limb sounding to measure vertical $CO_2$ profiles (e.g., Rinsland et al. 2010; Sioris et al. 2014).

Here we explore the detectability of $(N_2)_2$ in terrestrial planetary atmospheres using modern Earth as an exoplanet analog. In Section 2 we describe our models and the inclusion of $N_2$-$N_2$ and $N_2$-$O_2$ CIA absorption. We then report the detection of $(N_2)_2$ in Earth's disk-integrated spectrum using comparison data from *EPOXI* and perform a series of sensitivity tests to conclusively demonstrate the origin of the reported absorption. Additionally, we quantify the maximum transit transmission signal produced by $(N_2)_2$ for an Earth-Sun analog. In Section 3, we quantify the effect of $(N_2)_2$ in synthetic emission spectra of planets using self-consistent atmosphere-climate models. In Section 4, we simulate the transit transmission signal of $(N_2)_2$ for both a modern Earth analog and planets with $N_2$-$H_2$ atmospheres orbiting in the habitable zone (HZ) of a late type star. We discuss the implications of our results further in Section 5 and conclude in Section 6.

## 2. DETECTION OF $(N_2)_2$ IN EARTH'S DISK-AVERAGED SPECTRUM

### *2.1 The VPL 3D spectral Earth model and SMART*

To generate high-fidelity spectra to compare with spacecraft data, we use the Virtual Planetary Laboratory (VPL) three-dimensional, line-by-line, multiple scattering Earth model, which is fully described in Robinson et al. (2011). This model incorporates data from Earth-observing satellites to specify Earth's atmospheric and surface state, including snow, ice, and cloud cover. The model uses these factors as input to generate synthetic, time-dependent spectra of Earth, including absorption and scattering by the atmosphere, surface, and clouds and specular reflection by the ocean (Robinson et al. 2010). The VPL Earth model has been validated against broadband visible and NIR data from NASA's *EPOXI* mission, which repurposed the Deep Impact flyby spacecraft (Robinson et al. 2011; Livengood et al. 2011), and against phase-dependent, high-resolution visible and NIR spectral data from NASA's *LCROSS* mission (Robinson et al. 2014). The Spectral Mapping Atmospheric Radiative Transfer (SMART) model, developed by D. Crisp, is the core radiative transfer engine of the VPL spectral Earth model (Meadows & Crisp 1996; Crisp 1997). Below we share more details about the VPL spectral Earth model and SMART, which are used for spectral modeling throughout this work.

### *2.1.1 Earth model geometric calculations*

To generate a disk-integrated spectrum of the simulated planet, the VPL Earth model integrates the projected area weighted intensity of the planet in the direction of the observer (Robinson et al. 2014), which can be written as

$$F_\lambda(\hat{\boldsymbol{o}}, \hat{\boldsymbol{s}}) = \frac{{R_E}^2}{d^2} \int_{2\pi} I_\lambda(\hat{\boldsymbol{n}}, \hat{\boldsymbol{o}}, \hat{\boldsymbol{s}})(\hat{\boldsymbol{n}} \cdot \hat{\boldsymbol{o}}) d\omega \quad (1)$$

where $F_\lambda(\hat{o}, \hat{s})$ is the disk-integrated flux density measured by the observer, $R_E$ is the radius of the Earth (6378 km), $d$ is the distance from the planet to the observer, $I_\lambda(\hat{\boldsymbol{n}}, \hat{\boldsymbol{o}}, \hat{\boldsymbol{s}})$ is the location-dependent specific intensity directed at the observer, $\hat{\boldsymbol{o}}$ and $\hat{\boldsymbol{s}}$ are unit vectors in the direction of the observer and the Sun, respectively, $d\omega$ is an infinitesimal solid angle on the globe, and $\hat{\boldsymbol{n}}$ is a normal unit vector for the location of the planet's surface corresponding to $d\omega$. To calculate the disk-integrated spectrum, we use Equation (1) to integrate over the observable hemisphere (2π steradians).

The HEALpix parameterization (Gorski et al. 2005) is used to pixelate the atmosphere and surface and numerically evaluate Equation (1). If we divide the plane into N equal-area surface pixels (Robinson et al. 2014), Equation (1) can be written as

$$F_\lambda(\hat{\boldsymbol{o}}, \hat{\boldsymbol{s}}) = \frac{4\pi}{N}\frac{{R_E}^2}{d^2}\sum_{i\in O} I_\lambda(\hat{\boldsymbol{n}}_i, \hat{\boldsymbol{o}}, \hat{\boldsymbol{s}})(\hat{\boldsymbol{n}}_i \cdot \hat{\boldsymbol{o}}) \quad (2)$$

where $\hat{\boldsymbol{n}}_i$ is the unit vector describing the location on the sphere of surface pixel $i$, and $O$ is the set of indices of all observable pixels ($\widehat{\boldsymbol{n}_i} \cdot \hat{\boldsymbol{o}} > 0$). In this work we use 192 surface pixels of mixed type (coverage weighted averages of forest, grass, desert, ocean, and snow/ice), which are nested below 48 atmospheric pixels with 40 vertical layers. The temperature-pressure and gas mixing ratio profiles for each atmospheric pixel, the relative coverage of snow and clouds for each pixel, and the wind speed and direction used to calculate specular reflectance for each ocean pixel (Cox & Munk 1954) are determined by a suite of data from Earth-observing satellites (Robinson et al. 2011). Figure 1 shows the temperature and gas mixing ratio profiles for a representative mid-latitude ocean pixel (centroid: 41.8° S, 23.5° W). For each date the Earth model is run, SMART is used to generate a spectral database for every combination of atmospheric pixel (48), surface type (5), cloud type (4, plus clear sky), solar zenith angle (we use 0°, 15°, 30°, 45°, 60°, 75°, 80°, 85°, and 90°), observer zenith angle (we use the Gaussian angles 21.48°, 47.93°, 70.73°, and 86.01°), and observer azimuth angle (we use 0°, 30°, 60°, 90°, 120°, 150°, and 180°). The size of this spectral library is approximately 10,000 one-dimensional radiative transfer (SMART) runs, which takes approximately three weeks to generate using 100 CPUs. The model interpolates over this spectral library to calculate the spectrum of each surface pixel that is summed in Equation (2) by weighting according to the observer-planet-Sun viewing geometry and the relative coverage of each surface and cloud type for the pixel.

### *2.1.2 SMART description*

SMART is a one-dimensional, plane-parallel, line-by-line, multiple-scattering radiative transfer model. SMART takes as input vertical profiles of temperature, pressure, gas mixing ratios, and aerosol optical depths. With these inputs SMART calculates the monochromatic optical properties of each layer. Line absorption cross-sections are calculated with a program called Line-By-Line ABsorption Coefficients (LBLABC), also developed by D. Crisp, which determines the absorption coefficients of spectrally active gases given a line list database. For the simulations presented in this work, the HITRAN 2008 (Rothman et al. 2009) line lists were used to calculate the absorption coefficients for the spectrally absorbing monomer gases in our model ($O_2$, $H_2O$, $CO_2$, $O_3$, $N_2O$, $CH_4$ and CO). Throughout this paper, when calculating gaseous absorption coefficients with LBLABC, a line cutoff of 1000 $cm^{-1}$ is used (unless otherwise specified). LBLABC uses a Voigt line shape that includes Doppler broadening, Lorentzian (pressure) broadening, and Dicke (pressure) narrowing within a Voigt cutoff of 40 Doppler half-widths of the line-center (Meadows & Crisp 1996). At greater distances from the line center, a van Vleck-Weisskopf profile is used for all gases except for $H_2O$, $CO_2$, and $O_2$ (Goody & Yung 1989; pg. 100). For $H_2O$, the van Vleck-Weisskopf profile is altered at greater distances from line center by multiplication by an empirical wavenumber-dependent $\chi$ factor modified from Clough et al. (1989). For $CO_2$ a sub-Lorentzian profile is calculated based on a set of semi-empirical $\chi$ factors (Meadows & Crisp 1996). For $O_2$, a super-Lorentzian shape is calculated from Hirono & Nakazawa (1982) with an exponent of 1.958. LBLABC does not consider rotational broadening because it is negligible. The gas absorption coefficients determined by LBLABC are read in by SMART and multiplied by the gas mixing ratios and layer thicknesses to determine the normal-incidence gas extinction optical depths. The total extinction at each layer and each hyperfine spectral grid point is calculated by combining the gas extinction optical depths with those calculated for Rayleigh scattering and aerosols. We have updated SMART and the VPL Earth model to include $(N_2)_2$ and $O_2$-$N_2$ CIA absorption according to the empirical model presented in Lafferty et al. (1996). Below we briefly describe the inclusion of this absorption in our models.

### *2.2 Addition of $(N_2)_2$ and $N_2$-$O_2$ absorption to the Earth Model and SMART*

The $(N_2)_2$ and $N_2$-$O_2$ absorption included in our models is treated mathematically as collisionally-induced absorption (CIA). In general, the temperature- and wavenumber-dependent CIA absorption coefficients of a pure mixture of a collisionally absorbing gas A can be given as

$$\alpha(\nu, T, d_A) = B_{A-A}(\nu, T) d_A^2 \quad (3)$$

where $\alpha(\sigma, T, d_A)$ is the absorption coefficient (with units of $cm^{-1}$) for a given temperature (T), wavenumber ($\nu$), and density *($d_A$,* in units of amagat), and $B_{A-A}$ is a density-normalized CIA coefficient, usually given in $cm^{-1}$ $amagat^{-2}$. For a binary mixture of gases A and B, where B is spectrally inactive in the wavelength range of interest and is considered only as a collision partner to A, Equation (3) becomes

$$\alpha(\nu, T, d_A) = [d_A B_{A-A}(\nu, T) + d_B B_{A-B}(\nu, T)]d_A \quad (4)$$

Lafferty et al. (1996) present a simple empirical model to calculate the wavenumber and temperature-dependent normalized CIA coefficients for a mixture of pure $N_2$

$$B_{N_2-N_2}(\nu, T) = B_{N_2-N_2}{}^0(\nu)\exp\left[\beta_{N_2-N_2}{}^0(\nu)\left(\frac{1}{T_0} - \frac{1}{T}\right)\right] \quad (5)$$

where $B_{N_2-N_2}{}^0(\nu)$ and $\beta_{N_2-N_2}{}^0(\nu)$ are wavenumber-dependent empirical reference parameters given in Lafferty et al. (1996) (their Table 1), and $T_0$ is a reference temperature of 296 K. The empirically derived normalized CIA coefficients from this model are valid for a temperature range of 190-300 K. Figure 2 illustrates the normalized CIA coefficients (as a function of wavenumber or wavelength) calculated from Equation (5) for a variety of different temperatures. Lafferty et al. 1996 found the efficiency of $N_2$-$O_2$ absorption relative to $N_2$-$N_2$ absorption to be approximately independent of wavenumber, and given by

$$E^{N_2}_{O_2/N_2}(T) = \frac{B_{N_2-O_2}(\nu, T)}{B_{N_2-N_2}(\nu, T)} \cong \left[1.294 - 0.4545\left(\frac{T}{T_0}\right)\right] \quad (6)$$

Therefore, according to Equations (4, 5, and 6) the wavenumber- and temperature-dependent absorption coefficients due to CIA from a mixture of $N_2$ and $O_2$ gas can be calculated as

$$\alpha\left(\nu, T, d_{N_2}, d_{O_2}\right) = B_{N_2-N_2}(\nu, T)\left[d_{N_2} + E^{N_2}_{O_2/N_2}(T)d_{O_2}\right]d_{N_2} \quad (7)$$

Lafferty et al. (1996) give their absorption coefficients in wavenumbers. Note that we convert from wavenumber ($cm^{-1}$) to wavelength units (μm) for the model results presented throughout this work. We assume the CIA absorption in the 3.7-4.7 μm region is due entirely to $N_2$-$N_2$ and $N_2$-$O_2$ CIA, though CIA from other collision partners such as $N_2$-Ar will contribute a small, but insignificant, amount to the absorption, due to the small mixing ratios of these partners.

### *2.3 EPOXI Earth observations*

Our comparison observations consist of 1.0-4.5 μm disk-integrated NIR spectra taken of the Earth by the HRI instrument on the *Deep Impact* flyby spacecraft (Hampton et al. 2005; Klaasen et al. 2008) as part of the *EPOXI* mission. We present data model comparisons for four observations taken from 2008-March-18 through 2008-March-19 that span a full rotational cycle. These observations were made at a spacecraft to planet distance of 0.18 AU when Earth was at a phase angle of 57.7$^{\circ}$ (~77% illumination). Table 1 provides a summary of observations. Measured spectral irradiances were converted to distance-independent spectral radiances for data model comparison. The *EPOXI* Earth spectra are publicly available on the Planetary Data System (PDS) archive[1] and described more fully in Livengood et al. (2011).

### *2.4 Comparison of models and observations*

Simulated Earth model spectra were generated with a resolution of 1 cm$^{-1}$ and then degraded to the *EPOXI*/HRI resolution using a Gaussian convolution with FWHMs equal to the spectral bin size reported by the PDS for each spectral interval (for $\lambda$ = 4.15 μm, this is $\Delta\lambda \sim 0.013$ μm). Figure 3 compares NIR observations of the spectral radiance of the Earth with the synthetic Earth model data. The model cases where $(N_2)_2$ absorption is included provide a much better match to the data in the 4.0 – 4.15 μm range. The differences between the data and the model without collisional $(N_2)_2$ and $N_2$-$O_2$ absorption reach 31-40% at 4.15 μm, whereas the residuals between the data and the model with $(N_2)_2$ or with $(N_2)_2$ and $N_2$-$O_2$ absorption is less than 5%. The $(N_2)_2$ feature is robust and relatively constant (residuals within a few percent) through rotational phase, implying that the sensitivity of the spectral flux to surface features is small in this wavelength range.

[1] http://pds.nasa.gov/

There does remain a sharply peaked residual at the edge of the $CO_2$ band even when accounting for $(N_2)_2$ and $N_2$-$O_2$ absorption. The offset we observe between the EPOXI and Earth model spectra at the bottom of the 4.3 um $CO_2$ band, and the difference in line shape at the edge of the band, could be explained by non-LTE emission of $CO_2$ in the upper atmosphere, which is not accounted for in our model. Non-thermal emission by $CO_2$ near 4.3 μm is a well-studied phenomenon, seen in limb soundings of the atmospheres of Earth (e.g., López-Puertas & Taylor 1989), Venus (e.g, Gilli et al. 2009), and Mars (e.g., López-Valverde et al. 2005). Non-LTE models predict a peak radiance near 4.28 μm (~2337 $cm^{-1}$) (e.g., López-Valverde et al. 2011), which is the wavelength we observe the largest offset between our model spectra and the EPOXI data within the bottom of the band. Disk-integrated observations of Earth (such as the EPOXI observations presented here) include limb paths, where non-LTE contributions can be relatively large due to path length effects. Non-thermal $CO_2$ emission varies as a function of altitude, latitude, and solar zenith angle, and recent calculations indicate that non-LTE emission from $CO_2$ can account for the offset between our model and the data at the bottom of the 4.3 um $CO_2$ band (López-Valverde et al. 2011). Importantly, at wavelengths shortward of 4.15 μm where the effect of $(N_2)_2$ is observed, the atmospheric transmission is high and insensitive to the temperatures in the upper atmosphere. We conduct a series of sensitivity tests in Section 2.5 to show the peaked residuals or the absorption we attribute to $(N_2)_2$ and $N_2$-$O_2$ does not originate from simple differences between the width of the $CO_2$ band in the data and our model.

To our knowledge, this is the first time $N_2$ has been demonstrated in Earth's disk-averaged radiance spectrum. Spectra of Earth from other studies, including *Galileo*/NIMS spectra (Sagan et al. 1993; their figure 1c), *Mars Express*/OMEGA[2] IR spectra (Tinetti et al. 2006; their figure 10), and spatially resolved *ROSETTA*/VIRTIS spectra (Hurley et al. 2014; their figure 3) show a 4 um "roll off" on the short wavelength side of the $CO_2$ band with a similar spectral shape to the $(N_2)_2$ feature we present here. However, these studies do not attribute this feature to $(N_2)_2$. It should be noted that a disk-averaged *ROSETTA*/VIRTIS spectrum presented in Irwin et al. (2014; their figure 3) does not appear to show the same shape as seen in our Figure 2. However, the shape of their measured spectrum is also not consistent with the other studies mentioned above. Comparison synthetic spectra presented by Irwin et al. (2014) are also poor fits to their measured data in the 2.2-5 μm spectral range, except for the core of the 4.3 μm $CO_2$ band.

### *2.5 Sensitivity tests*

To demonstrate that the spectral "roll off" seen near 4 μm is due to collisional $(N_2)_2$ absorption and not an artifact or the influence of other gaseous absorption, we conduct a series of sensitivity tests. In our first test, we examine the sensitivity of the residuals between the *EPOXI* spectral data and the synthetic Earth model comparisons to the assumed FWHM used to degrade the high-resolution synthetic spectra. As stated in Section 2.4, we degraded our synthetic spectra to the same spectral resolution as the *EPOXI* measurements using a Gaussian convolution with FWHMs equal to the spectral bin size reported in the PDS for each wavelength. We determine the effect of the assumed FWHM on the shape of the data-model residuals by degrading one set of synthetic spectra (top left panel in Figure 3; sub-spacecraft longitude of 215° W) with 1, 2, and 4 times the FWHM at 4.15 μm (0.013 μm). Figure 4 shows the result. While a FWHM of 0.052 μm (4 × 0.013 μm) decreases the residuals somewhat shortward of 4.16 μm, it leads to an increase in the residuals longward of 4.2 μm and an overall poorer fit to the spectral shape.

---

[2] http://sci.esa.int/mars-express/31033-objectives/?fbodylongid=661

Next we use a series of one-dimensional SMART runs to compare the absorption of $(N_2)_2$ and $N_2$-$O_2$ to absorption by $CO_2$ and $N_2O$, which are also spectrally active in this wavelength region. Besides helping to demonstrate the detection of $(N_2)_2$ in Earth's spectrum, these sensitivity tests are also useful guides to the robustness of $(N_2)_2$ absorption in atmospheres with slightly different compositions. For each of these SMART runs, we use a temperature and gas mixing ratio profile from one of the retrieved profiles used in the VPL Earth Model (Figure 2). We assume the surface is an ocean (McLinden et al. 1997) with no clouds, a surface temperature of $T_{surf}$ = 288.2 K, a solar zenith angle of 60° and an azimuth angle of 0°. Figure 5 illustrates the results of these tests. In Figure 5a we show one-dimensional clear sky SMART spectra with and without $(N_2)_2$ and $N_2$-$O_2$ absorption, which looks similar to comparisons of the data and model without $(N_2)_2$ and $N_2$-$O_2$ absorption in Figure 3 (though the one-dimensional SMART spectra are not directly comparable to the three-dimensional disk-integrated Earth model spectra). In the remaining tests we do not include collisional $N_2$ absorption. In Figure 5b we show the effect of doubling or halving the $CO_2$ mixing ratio, demonstrating this saturated band is relatively insensitive to differences in $CO_2$. In Figure 5c we experiment with different line cutoffs used in LBLABC to generate our absorption coefficients (see Section 2.1.2). This panel shows that line cutoffs larger than 100 $cm^{-1}$ would be adequate to model the 4.3 μm $CO_2$ band. For all other cases we used line cutoffs of 1000 $cm^{-1}$, which is larger than the width of the plot window (a $\Delta\nu$ of 1000 $cm^{-1}$ is a $\Delta\lambda$ of ~1.76 μm at $\lambda$ = 4.2 μm). Finally in Figure 5d we show the effect of doubling and halving the $N_2O$ concentration. While $N_2O$ absorbs between 4.0-4.2 μm, its effect is negligible. In each of the panels of Figure 5, the *EPOXI* spectrum of the Earth with a sub-spacecraft longitude of 215° W (see Figure 3 and Table 1) is added for contrast. Note that the EPOXI spectrum of the disk-integrated Earth and the 1-D SMART synthetic spectra of a single sounding are not directly comparable. This is because the observed Earth contains clouds, heterogeneous surfaces including land as well as ocean, and a distribution of solar zenith and observer angles, which are not present in the simple synthetic spectra - with an ocean surface and no clouds - generated to show sensitivity to a given atmospheric constituent or line cutoff assumption (see Figure 3 for the directly comparable synthetic spectra).

### *2.6 Simulated transmission spectrum of the Earth-Sun with $(N_2)_2$ absorption*

To investigate the detectability of $(N_2)_2$ in Earth's transmission spectrum and inform future observations of Earth-Sun analogs, we used the VPL one-dimensional transmission model, which includes the effects of refraction and has been validated for Earth observations (Misra et al. 2014b). We assumed the same temperature and mixing ratio profiles as given in Figure 1 (including absorbing gases $O_2$, $H_2O$, $CO_2$, $O_3$, $N_2O$, $CH_4$ and CO) and no clouds or aerosols. The spectral resolution for the simulated transit transmission spectrum is $\Delta\lambda = 0.005$ μm. Due to refraction only the top ~0.2 bars of the atmosphere can be probed for a distant observer with the Earth-Sun geometry (García Muñoz et al. 2012; Bétrémieux & Kaltenegger 2013). Figure 6 shows the largest effect $(N_2)_2$ could have on the transmission spectrum of an Earth-Sun analog. The maximum signal is < 0.1 ppm due to the limiting tangent pressure (defined as the pressure above which only 50% or less of light is transmitted) of < ~0.2 bar and the strong density dependence of collisional pair and/or dimer absorption (Misra et al. 2014a). While $N_2$ absorption would be undetectable in an Earth-Sun analog due to high limiting tangent altitudes from refraction, circumstances are more favorable for planets orbiting within the habitable zone of lower mass stars where refraction allows lower altitudes to be probed (Bétrémieux & Kaltenegger 2014; Misra et al. 2014b). We investigate this further in Section 4.

## 3. RADIANCE SPECTRA OF $N_2$-DOMINATED ATMOSPHERES

### *3.1 Self-consistent $N_2$-$CO_2$-$H_2O$ Atmospheres*

To explore the general case of $(N_2)_2$ absorption in the emission spectra of terrestrial planets, we used a modified version of a publicly available[3] radiative-convective model (Kasting et al. 1984; Segura et al. 2005; Haqq-Misra et al. 2008) to construct self-consistent temperature-pressure and water vapor mixing ratio profiles of pure $N_2$-$CO_2$-$H_2O$ atmospheres to use as input for our one-dimensional radiative transfer model (SMART). Atmospheres that have $CO_2$ and $H_2O$ concentrations consistent with their surface temperatures are useful because of the possible degeneracy between $CO_2$ absorption and $(N_2)_2$ absorption in the wings of the 4.3 μm $CO_2$ band. In contrast, we could have explored cases with increased $N_2$ but with constant $CO_2$, while maintaining the same surface temperature (which sets the magnitude of the thermal flux), but this would be unrealistic because the planet's $CO_2$ concentration would be inconsistent with its temperature and thermal flux. Pure $N_2$-$CO_2$-$H_2O$ atmospheres are chosen to provide terrestrial planet base cases without complicating assumptions about the origin of major or trace gases that may be due to life (such as $O_2$ from photosynthetic organisms), and the development of trace gases from photochemistry that may also be greenhouse absorbers (such as $O_3$ from $O_2$). The gray surface Bond albedo of the planet was chosen such that in an Earth-like scenario with modern (pre-industrial) mixing ratios of $CO_2$, $H_2O$, $O_3$, $N_2O$, $N_2$, and $CH_4$ the surface temperature converged to 288 K. This albedo ($A_B$ =0.238) was used for each surface pressure and temperature scenario. This is effectively a tuning parameter that allows a one-dimensional climate model to match Earth's modern surface temperature for calibration (Kasting et al. 1993).

[3] http://vpl.astro.washington.edu/sci/AntiModels/models09.html

We assumed the same incident solar flux as Earth currently receives and adjusted the $CO_2$ concentration until the surface temperature of the planet converged to the desired value. The water vapor mixing ratios were calculated using a relative humidity profile with a surface relative humidity of 80% (Manabe & Wetherald 1967). The $CO_2$ mixing ratio profile was constant with altitude. $N_2$ constituted the remaining atmospheric volume of each layer after $CO_2$ and $H_2O$ were accounted for. Six pressure scenarios were considered, with $P_{surf}$ = 0.2, 0.5, 1.0, 2.0, 5.0, and 10.0 bars. Three temperature scenarios were calculated with a solar input spectrum: 273 K, 288 K, and 300 K. A similar set of atmospheric profiles were calculated for a 288 K surface temperature planet with the incident spectrum of AD Leo, an M3.5 dwarf with an effective temperature of 3390 K (Rojas-Ayala et al. 2012). We used the same AD Leo spectrum as in Segura et al. (2005, 2010), which was scaled from the Earth-Sun flux equivalent distance by an albedo correction factor ($A_{fac}$) of 0.9 to correct for the absorption of redder stellar radiation by an Earth-like atmosphere. This distance from the star to the planet was calculated as r = 1 AU $[(L/L_\odot)/A_{fac}]^{1/2}$, following Segura et al. (2005). This produced a distance of r = 0.16 AU and a normalized flux of 1225 W/m$^2$ for the AD Leo cases, compared to an incident flux of 1360 W/m$^2$ for the Earth-Sun scenarios. AD Leo was chosen to represent the category of stars with effective temperatures less than that of the Sun, where a larger number of habitable zone planets may be found (Dressing & Charbonneau 2013; Kopparapu 2013). AD Leo is often used in the study of the atmospheres of planets orbiting different spectral classes of star because a complete, calibrated spectrum of this star is available from the UV to the MIR (Domagal-Goldman et al. 2014; von Paris et al. 2013a; Segura et al. 2010).

To study the potential for $CO_2$-dominated (i.e., Mars-like) atmospheres to mimic the $(N_2)_2$ feature, a set of temperature-pressure and gas mixing ratio profiles were calculated using the same radiative-convective model as described above, but for atmospheres dominated by $CO_2$. This allows us to compare the change in band intensity and shape for self-consistent atmospheres with $CO_2$ (broadened by $N_2$) plus $(N_2)_2$, and self-consistent atmospheres with only $CO_2$. For these atmospheres, the volume fraction of $CO_2$ was fixed at 96%, equivalent to Mars' $CO_2$ mixing ratio, with small amounts of $N_2$ constituting the remaining volume after accounting for $H_2O$. Instead of using the $CO_2$ fraction ($f_{CO2}$) as the dial to converge the model to the desired surface temperatures, the surface pressure was adjusted.

The properties of each atmosphere are given in Tables 2 ($N_2$-$CO_2$-$H_2O$ atmospheres) and Table 3 ($CO_2$-dominated atmospheres) and their temperature-pressure profiles are shown in Figure 7. There are two competing factors that set the $CO_2$ mixing ratio. As the surface pressure (equivalently atmospheric mass for constant surface gravity and temperature) increases, the total amount of $CO_2$ (and $H_2O$) also increases, forcing the $CO_2$ mole fraction to be tuned downwards. However, as the atmospheric mass increases, the planetary albedo also increases because of enhanced Rayleigh scattering to space. This cools the surface (by reducing the amount of incident solar radiation) and necessitates more $CO_2$. (It should also be noted that $CO_2$ is also much more efficient at Rayleigh scattering than $N_2$ and for this reason, along with the saturation of $CO_2$ absorption bands, there are diminishing returns for adding more $CO_2$). The temperature-pressure profiles are affected by the mole fraction of $CO_2$ in the upper atmosphere. More $CO_2$ enhances the radiative cooling in these upper layers, and because the $CO_2$ mixing ratio is constant with altitude for each scenario, this results in the radiative cooling in the upper atmosphere being more effective for the atmospheres with lower total surface pressure and thus larger $CO_2$ mixing ratios. It is important to note that because the absorption by $(N_2)_2$ is highly density dependent and weakly temperature dependent, our results are insensitive to the temperature profiles of the upper atmosphere. Segura et al. (2005) showed that a lower integrated stellar flux is required from M dwarfs to achieve the same planetary surface temperature, due to the lower planetary albedo caused by gaseous absorption of more plentiful incoming IR radiation. We follow the convention suggested by Segura et al. (2005), and use a lower integrated stellar flux for our AD Leo planets. However, the pressures and compositions of our atmospheres differed from Segura et al. (2005). As a result, more $CO_2$ was required for the $P_{surf} \leq 1$ bar AD Leo cases to achieve the same surface temperature ($T_{surf}$=288 K) as the Earth-Sun cases, as these cases have less efficient albedo augmentation due to the lack of absorbing gases other than $H_2O$ and $CO_2$. For the AD Leo cases with $P_{surf} \geq 2$ bar, the combination of the redder spectrum of AD Leo and the greater total $H_2O$ content of the atmospheres (see Table 2), necessitated lower $CO_2$ fractions than for the equivalent Earth-Sun cases. Note that for a surface pressure of $P \leq 0.2$ bars, a surface temperature of 300 K assuming a solar spectrum and a surface temperature of 288 K with an AD Leo spectrum are inconsistent with even a pure $CO_2$ atmosphere, so we have not included those scenarios in Figure 7, Table 2, or subsequent figures.

### *3.2 Simulated Radiance Spectra*

For each of the self-consistent $N_2$-$CO_2$-$H_2O$ temperature pressure and mixing ratio profiles described in section 3.1, we generated spectra using SMART. One series of spectra assumed only thermal emission, shown in Figure 8. In these cases, the spectral albedo in the 3.4-5.0 μm range was set to zero and the emissivity was set to 1 to isolate thermal emission, because reflected stellar light contributes much less than thermal emission to the total spectral flux from the planet, but will depend on the surface spectral albedo and cloud emissivities and altitudes. The albedo in this wavelength range contributes very little to the Bond albedo of the planet, so this assumption is not inconsistent with Section 3.1. The resulting spectra demonstrate strong dependence on $N_2$ abundance for surface pressures exceeding 1 bar with the flux being reduced in the wings of the 4.3 μm $CO_2$ band by up to 100% for sufficiently high $N_2$ abundances. The results from our one-dimensional models are consistent with the *EPOXI*-VPL Earth model comparison. For the $P_{surf}$ =1 bar, $T_{surf}$ = 288 K synthetic spectrum from the one dimensional model, the $(N_2)_2$ absorption has a maximum effect of 35% on the spectral flux, which is similar to that observed by *EPOXI* for Earth's disk-averaged spectrum. The effect of $(N_2)_2$ absorption decreases for self-consistent scenarios with higher surface temperatures. This is because the higher temperature planets require more $CO_2$ to warm their surfaces, and thus the width of the 4.3 μm $CO_2$ band increases and masks $(N_2)_2$ absorption.

We generated another series of spectra that include reflected light, shown in Figure 9. In these cases, the spectral albedo of the planet was set to 10% for the 3.4-5.0 μm wavelength range. The spectrum used to model the incident light from the Sun is a Kurucz model (Kurucz 1995), and the spectrum used for AD Leo is the same used in Section 3.1 (Segura et al. 2005, 2010). The increase in the total spectral flux is especially notable for AD Leo, whose SED is shifted more strongly into the NIR compared to the Sun due to its lower effective temperature. The peak relative $(N_2)_2$ contribution is similar for the scenarios that include reflected light (Figure 9) and the scenarios that do not include reflected light (Figure 8), though the differences for the cases with reflected light are greater at shorter wavelengths where there is more incident flux from the star. We do not attempt the inverse problem here, but note that accounting for $(N_2)_2$ absorption could potentially provide leverage to constrain $N_2$ abundances with future retrieval algorithms when applied to direct-imaging spectra of planets with substantial atmospheres (e.g., Benneke & Seager 2012; Line et al. 2013; von Paris et al. 2013b).

## 4. TRANSMISSION SPECTRA OF PLANETS WITH $N_2$ ATMOSPHERES ORBITING LATE TYPE STARS

Transit transmission observations will be more favorable for planets orbiting the HZ of late type stars, where the lower atmosphere may be probed because geometric refraction limits approach pressures greater than (or equivalent altitudes lower than) ~1 bar for $N_2$-dominated atmospheres (Bétrémieux & Kaltenegger 2014; Misra et al. 2014b). Additionally, JWST may have the opportunity to characterize one or more super-Earths in the habitable zones of late type stars (Deming et al. 2009), increasing the importance of studying these cases. We generated a synthetic transit transmission spectrum of an Earth orbiting in the HZ of an M5V star (a = 0.05 AU, R = 0.2 $R_\odot$) using the model described in Misra et al. (2014b). The conservative and optimistic inner edge limits for the habitable zone of a $T_{eff}$ = 2800 K, L = 0.0022 $L_\odot$ M5V star are a = 0.049 AU and a = 0.039 AU, respectively (Kopparapu et al. 2013)[4]. We assumed the same set of temperature and gas mixing ratios in Figure 1. Figure 10 shows the effect of $(N_2)_2$ in this test case. The maximum spectral transmission depth is ~3 ppm, which is almost two orders of magnitude greater than that seen for the Earth-Sun analog (Figure 6). This difference is primarily due to the smaller radius of the M5V host star relative to that of the Sun. An additional contribution comes from the combination of refraction - which permits deeper altitudes to be probed for a planet orbiting in the habitable zone of an M5V star - and the strong density dependence of $(N_2)_2$ absorption. However, this is still less than the anticipated noise level for JWST/NIRSPEC observations of an M5V host star at 4.1 μm, which Misra et al. (2014b) calculate as ~5 ppm (see their Figure 13) assuming all transits in a five year period are observed.

[4] See the VPL HZ calculator based on Kopparapu et al. (2013): http://depts.washington.edu/naivpl/content/hz-calculator

Figure 10 does not represent the most ideal scenario for detecting $(N_2)_2$. The spectral impact of $(N_2)_2$ is also function of the atmospheric scale height, the mean molecular weight of the atmosphere, and the concentrations of $CO_2$ and $N_2O$, since both gases have overlapping absorption features. To demonstrate the maximum plausible detectability of $(N_2)_2$ with transit transmission spectroscopy, we calculate transmission spectra of atmospheres dominated by mixtures of $N_2$ and $H_2$ with different trace fractions of $CO_2$ and $N_2O$. The addition of a low molecular weight component ($H_2$) will increase the scale height and consequently the transit height variations (or, equivalently, spectral transit depths) due to the spectrally absorbing gases (Miller-Ricci et al. 2009). Below we describe the model atmospheres and the synthetic transmission spectra we generate.

*4.1 Analytic $N_2$-$H_2$ Atmospheres*

We employed a simple model for calculating the temperature-pressure-altitude structure of an atmosphere that consists of pure mixtures of one or more gases. We assumed a surface pressure of 1 bar, a surface temperature of $T_{surf}$ = 288 K, and an isothermal stratospheric temperature of $T_{strat}$ = 200 K, which is used when simulating atmospheres without stratospheric UV absorbers (Schindler & Kasting 2000; Haqq-Misra et al. 2008; Domagal-Goldman et al. 2011). We used the thin layer approximation, so the gravity remained constant with altitude. The altitude grid extended from 0 km to 1000 km in intervals that increased with altitude from 3-100 km, though we found our results were insensitive to the altitude intervals used. The temperature changed at each altitude below the stratosphere based on the lapse rate. The lapse rate of the atmosphere was calculated as:

$$L = \frac{dT}{dz} = -\frac{g}{c_p} \quad (8)$$

where L is the dry adiabatic lapse rate, g is the surface gravity, and $c_p$ is the heat capacity of the gas mixture at constant pressure. The heat capacity is a weak function of temperature, which we ignore in this simple treatment. The temperature at each altitude level below the tropopause is calculated as:

$$T = T_{surf} + L * Z \quad (9)$$

where Z is the altitude point in the array and $T_{surf}$ is the input surface temperature. For all values of $T < T_{strat}$ the temperature is set to $T_{strat}$. To calculate the pressure below the stratosphere, the barometric equation is used:

$$P = P_{surf}\left(\frac{T_{surf}}{T}\right)^{\left(\frac{g}{RL}\right)} \text{ for } T > T_{strat} \quad (10)$$

Here, P is the pressure at a grid point, $P_{surf}$ is the surface pressure, and R is the gas constant of the mixture. The pressure above the troposphere is calculated as:

$$P = P_{strat} * e^{-\left[\frac{(Z-Z_{strat})}{H_{strat}}\right]} \quad (11)$$

where $Z_{strat}$ is the altitude where the temperature lapses to the stratospheric temperature, $P_{strat}$ is the pressure at that altitude as calculated by (10), and $H_{strat}$ is:

$$H_{strat} = \frac{k_B T_{strat}}{\mu g} \quad (12)$$

where $k_B$ is the Boltzmann constant, and μ is the mean molecular mass of the atmosphere. The properties μ ($N_2$=28.02 $\frac{g}{mol}$, $H_2$=2.06 $\frac{g}{mol}$), $c_p$ ($N_2$=1040 $\frac{J}{kg \cdot K}$, $H_2$=14320 $\frac{J}{kg \cdot K}$), and R ($N_2$=296.8 $\frac{J}{kg \cdot K}$, $H_2$= 4124 $\frac{J}{kg \cdot K}$) are given in terms of mass quantities for individual gases; we therefore weight according to mass mixing ratio when calculating these variables. We used the above equations to calculate the temperature-pressure-altitude structure for gas mixtures of 100% $N_2$, 75% $N_2$, 50% $N_2$, and 0% $N_2$ with the remaining percentages consisting of $H_2$.

### *4.2 Transmission Spectra of $N_2$-$H_2$ atmospheres*

We modeled the transmission spectra of the analytic $N_2$-$H_2$ atmospheres described above, including $(N_2)_2$ absorption (Lafferty et al. 1996). Since the $(N_2)_2$ feature substantially overlaps with the 4.3 μm $CO_2$ band, we modeled scenarios including different concentrations of $CO_2$ (Rothman et al. 2009): 0 ppm, 1 ppm, 10 ppm, 100 ppm, and 1000 ppm. Figure 11 shows the effect of $(N_2)_2$ in the transmission spectrum for the cases described from 3.5 – 5.0 μm. The only sources of opacity for these simple synthetic spectra are $CO_2$ and $(N_2)_2$ absorption. We find that with a 100% $N_2$ atmosphere, the maximum spectral transit signal produced for the Earth-sized planet is 10 ppm, while the 25% $N_2$/ 75% $H_2$ case produces a spectral transmission signal of about 50 ppm at 4.3 μm (with no $CO_2$). For the 50% $N_2$/50% $H_2$ cases with trace $CO_2$ concentrations, the peak spectral transmission depths due to $(N_2)_2$ range from 20 ppm (0 ppm $CO_2$) to 8 ppm (1000 ppm $CO_2$). The transit depth variations would be smaller for super-Earth atmospheres containing the same mixing ratios of $N_2$, $H_2$, and $CO_2$ due to a smaller scale height, though super-Earths are more likely to retain larger fractions of $H_2$ (Miller-Ricci et al. 2009).

Because $N_2O$ also absorbs in this wavelength range (see Figure 5d), we illustrate the sensitivity of $(N_2)_2$ absorption in transit transmission to different $N_2O$ abundances. The introduction of $N_2O$ would provide opacity at these wavelengths preventing an observer from probing the lower altitudes where $(N_2)_2$ can most impact the spectrum. We generate transmission spectra for a 75% $N_2$, 25% $H_2$ atmosphere with 100 ppm $CO_2$ and 0 ppb, 10 ppb, 100 ppb, 1000 ppb, and 10000 ppb $N_2O$ (Figure 12). For comparison, the pre-industrial concentration of $N_2O$ in Earth's atmosphere is 265 ppb (Flückiger et al. 2002) and the atmospheric concentration of this biogenic gas likely varied greatly on Earth over geologic time (Buick 2007). With 100 ppm $CO_2$, 75% $N_2$, 25% $H_2$ and trace $N_2O$, peak spectral transmission depth variations due uniquely to $(N_2)_2$ range from just under 10 ppm (0 ppb $N_2O$) to just under 6 ppm (10000 ppb $N_2O$). The integrated $\Delta(N_2)_2$ residuals of the spectral transmission depths from 3.7 μm to 4.8 μm decrease more strongly with increased $N_2O$ mixing ratios than the peak spectral transmission depths.

## 5. DISCUSSION

### *5.1 Detecting bulk gases to establish habitability*

We have shown that the absolute flux differences at the relevant wavelengths for $(N_2)_2$ absorption are significant (~30-35%) for Earth-analog cases, potentially allowing for the detection of $N_2$ in exoplanet atmospheres. Characterizing the bulk atmosphere of a terrestrial exoplanet will allow for the determination of the atmospheric mass, and therefore the surface pressure if the surface gravity is known. Liquid water stability, the common definition of habitability, depends on both temperature and pressure. To establish habitability, it is not necessary to quantify the pressure exactly, but to constrain it to be above the stability criterion for water within a known temperature range for the planet. We have demonstrated that $(N_2)_2$ would not create a measurable effect at the low surface pressures that preclude surface water stability for habitable temperatures.

### *5.2 Biosignature confirmation with $(N_2)_2$*

Detection of $(N_2)_2$ could also help discriminate whether the presence of $O_2$ is more or less likely to be due to life. It has been postulated that the abiotic buildup of $O_2$ could occur on planets with a limited abundance of non-condensable gases (Wordsworth & Pierrehumbert 2014) that would allow the atmosphere of those planets to become $H_2O$-dominated. For each possible surface temperature a minimum non-condensable gas partial pressure is required to generate a cold trap and stabilize an atmosphere against rapid H loss through $H_2O$ photolysis in the stratosphere and consequent $O_2$ buildup. An $N_2$ surface pressure of above a few percent of modern day Earth's is required to prevent the upper atmosphere of a planet with $T_{surf}$ = 288 K from becoming rich in $H_2O$ (Wordsworth & Pierrehumbert 2014). We find that the effect of $(N_2)_2$ on the spectral flux does not become significant (exceeding 5%) in emission spectra unless the surface pressure is greater than 0.5 bar. Therefore, direct detection of $(N_2)_2$ would imply a non-condensable gas partial pressure larger than that required to rule out this abiotic $O_2$ scenario and bolster the case that detected $O_2$ is biological in origin. While H loss due the lack of a cold trap is not the only possible source of abiotic $O_2$ in a terrestrial planet atmosphere (e.g., Domagal-Goldman et al. 2014; Luger & Barnes 2015), it is the only mechanism that has thus far been proposed that could operate at all locations of the habitable zone for all spectral classes of host star.

### *5.3 Detecting $(N_2)_2$ in transit transmission*

We have demonstrated that $(N_2)_2$ absorption can create spectral transmission depths of about 50 ppm for $N_2$-$H_2$ atmospheres without $CO_2$ and 10 ppm for $N_2$ atmospheres without $CO_2$. The detection of $(N_2)_2$ in the transit transmission spectrum of an $N_2$-doimated atmosphere would require an atmosphere to be predominately cloud-free down to a ~0.5 bar pressure level. The requirement to probe deep altitudes is potentially problematic because long path lengths through an atmosphere in transit transmission mean that even tenuous clouds or hazes can be optically thick at high altitudes (Fortney 2005). The flat transmission spectra of GJ 1214b (Kreidberg et al. 2014) and GJ 436b (Knutson et al. 2014) demonstrate that high-altitude clouds might be commonplace in close-in exoplanets. However, NIR transmission spectrum of the "warm Neptune" HAT-P-11 from WFC3 on *Hubble* demonstrates the presence of water vapor and indicates the planet's atmosphere is cloud-free at least down to the 1 mbar level and perhaps to levels where pressure are larger than 1 bar (Fraine et al. 2014). Thus, at least some planetary atmospheres may be characterizable at the pressure altitudes where $N_2$ collisional absorption would become detectable.

## 6. CONCLUSIONS

Although the $N_2$ molecule is extremely difficult to detect in planetary spectra, we have shown that $(N_2)_2$ is detectable in disk-integrated observations of the Earth from interplanetary spacecraft, creating a ~35% effect on the spectrum of the Earth at 4.15 μm. We explored the strength of the $(N_2)_2$ collisional pair on the emission spectra of self-consistent $N_2$-$H_2O$-$CO_2$ atmospheres and find a strong spectral response for $N_2$ abundances greater than 0.5 bar. A detectable amount of $N_2$ would imply surface pressures high enough to support liquid water and rule out the production of abiotic $O_2$ through one proposed mechanism. We find that $(N_2)_2$ absorption could produce transmission depths of 8 ppm at 4.1 μm in the spectra of planets orbiting in the habitable zones of late type stars with $N_2$-dominated atmospheres and transit depths several factors larger for $N_2$-$H_2$ atmospheres with larger scale heights. We demonstrate that for both direct imaging and transit transmission spectroscopy, $(N_2)_2$ could significantly widen the width of the 4.3 μm $CO_2$ band in thick $N_2$ atmospheres. The detection of $N_2$ would provide critical contextual information about planetary atmospheres and may be relevant to future exoplanet observing missions.

**Acknowledgements**

This work was supported by the NASA Astrobiology Institute's Virtual Planetary Laboratory Lead Team, funded through the NASA Astrobiology Institute under solicitation NNH12ZDA002C and Cooperative Agreement Number NNA13AA93A. T.R. gratefully acknowledges support from an appointment to the NASA Postdoctoral Program at NASA Ames Research Center, administered by Oak Ridge Affiliated Universities. This work benefited from the use of advanced computational, storage, and networking infrastructure provided by the Hyak supercomputer system at the University of Washington. This research has made use of NASA's Astrophysics Data System. We thank Dave Crisp and Robin Wordsworth for helpful comments and discussion. We are grateful for the comments of an anonymous reviewer, which allowed us to greatly improve this manuscript.

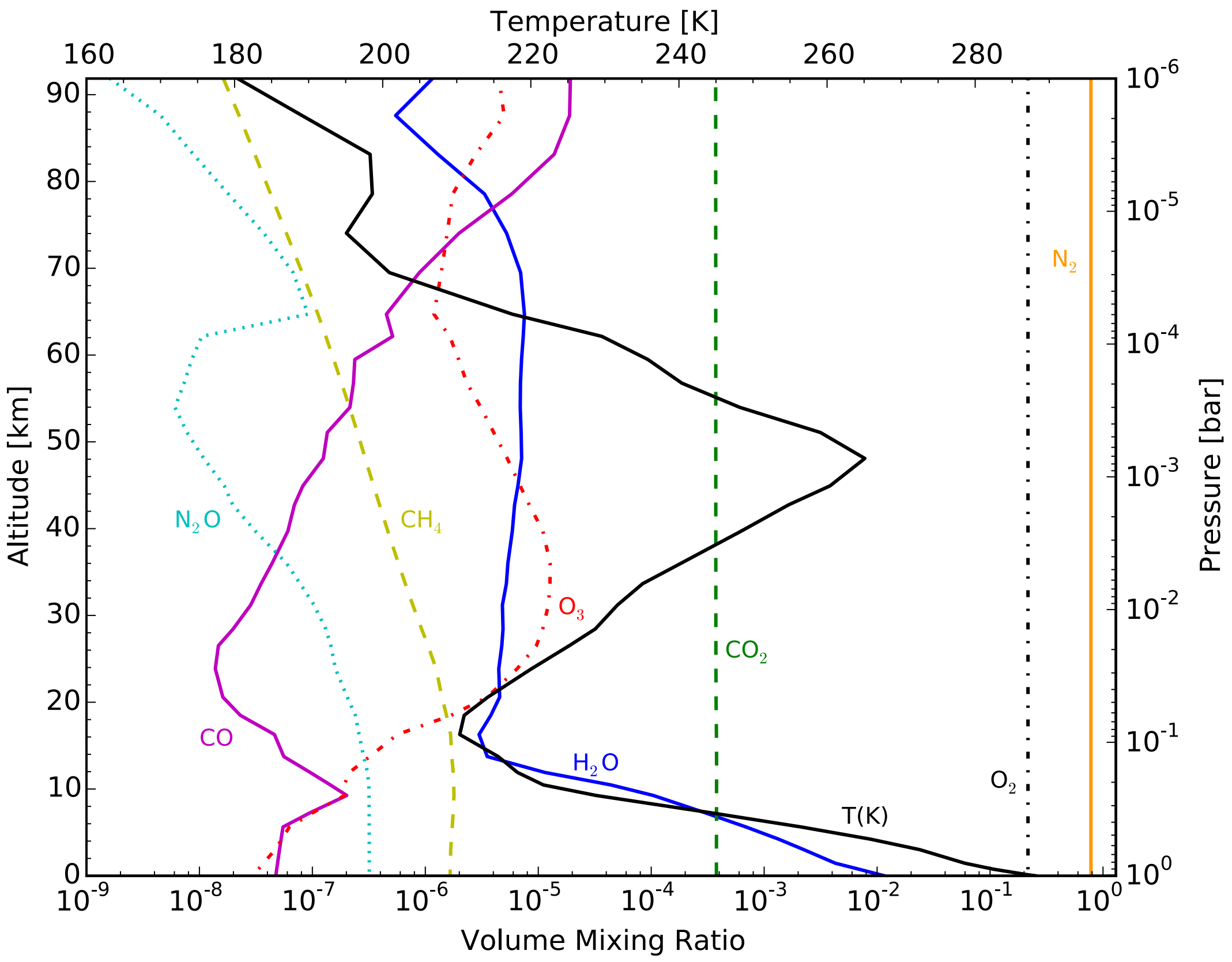

**Figure 1.** Temperature and gas volume mixing ratio profiles from a representative mid-latitude pixel (41.8° S, 23.5° W) from the VPL Earth model calculated for March 19, 2008. This model atmosphere was used for generating spectra for the sensitivity tests in Section 2.5 and the transit transmission spectra in Section 2.6 and Section 4.

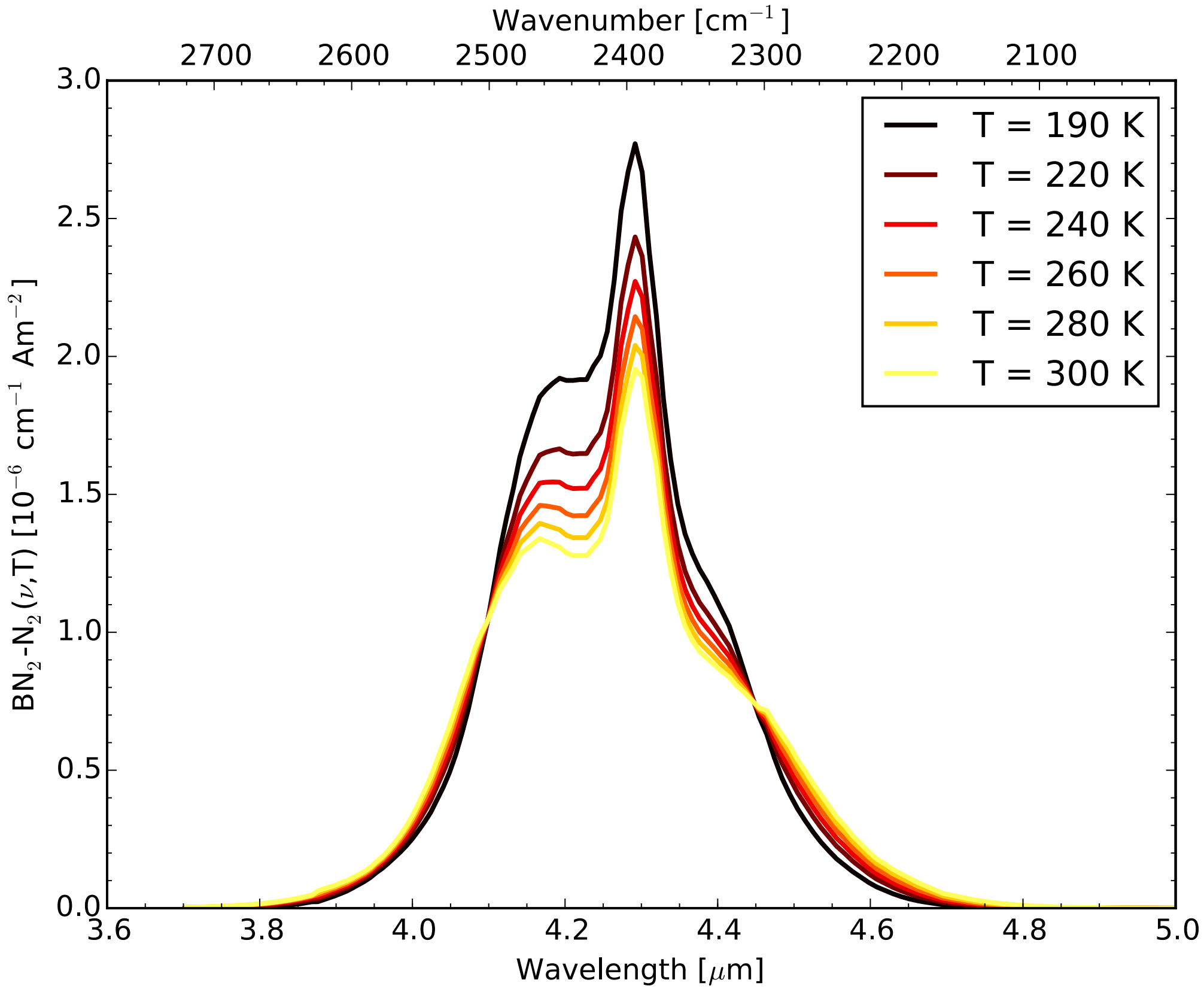

**Figure 2.** Density-normalized wavelength- and temperature-dependent $(N_2)_2$ absorption coefficients. The normalized absorption coefficients are calculated from Equation (5) and reference parameters given in Table 1 of Lafferty et al. (1996) for a gas of temperature T = 190 K (black), 220 K (dark red), 240 K (red), 260 K (dark orange), 280 K (light orange), and 300 K (yellow). The wavenumber (or wavelength) and temperature dependent absorption coefficients of a mixture of $N_2$ and $O_2$ gas can be calculated using Equation (7) given in the text.

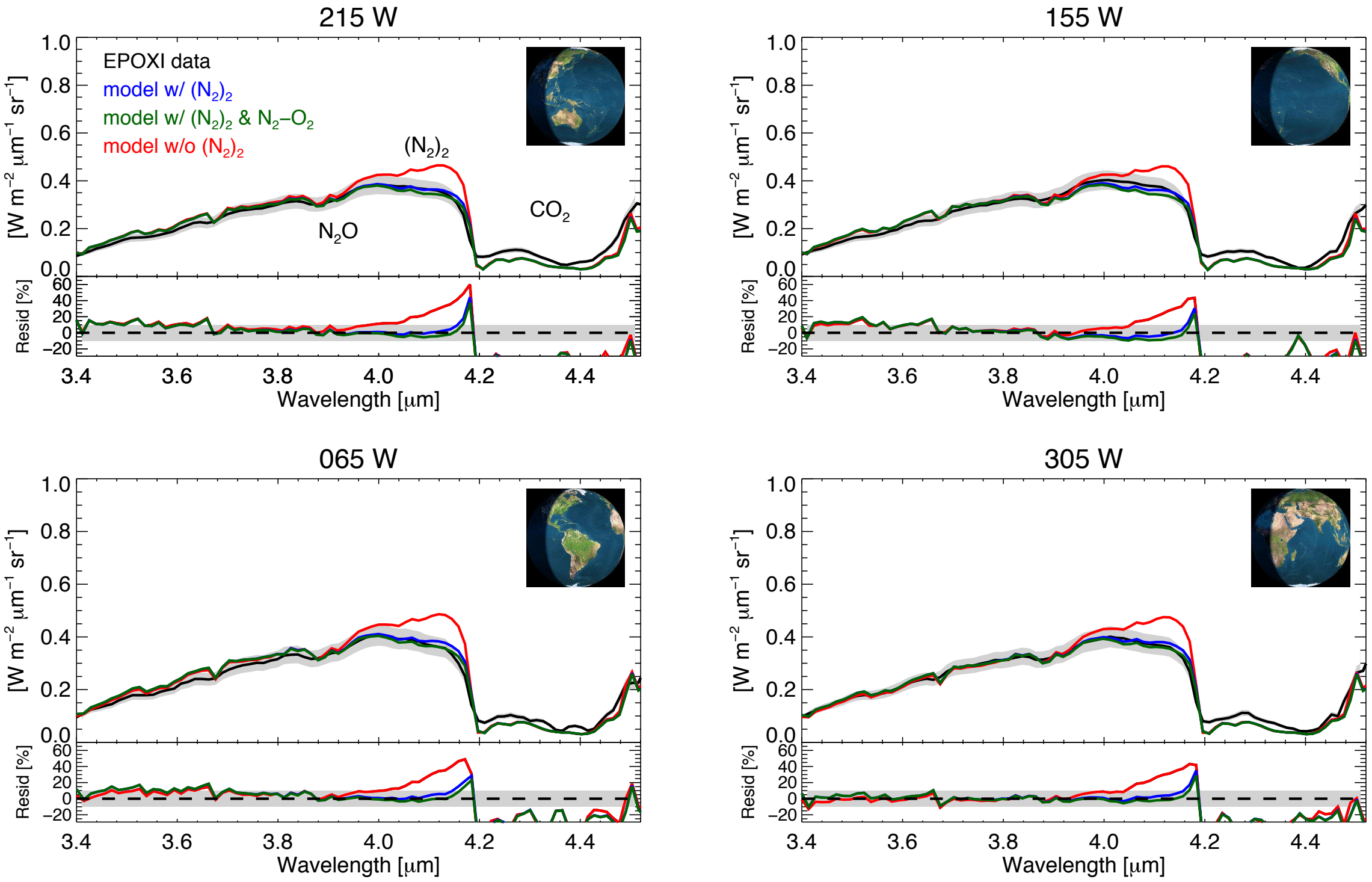

**Figure 3.** Detection of $(N_2)_2$ absorption in Earth's NIR spectra. Spectral radiances of the Earth as measured by the *EPOXI* NIR spectrograph (black), as generated by the VPL three-dimensional Earth Model without $(N_2)_2$ absorption (red), with $(N_2)_2$ absorption (blue), and with both $(N_2)_2$ and $N_2$-$O_2$ absorption (green). Titles provide the sub-spacecraft longitudes (also given in Table 1). The gray band is the calibration uncertainty for the NIR data (Klaasen et al. 2008). *Insets:* Illustrations of Earth's illumination as seen by the *EPOXI* spacecraft during each observation. Insets were generated with the Earth Moon Viewer originally developed by J. Walker (http://www.fourmilab.ch/earthview/).

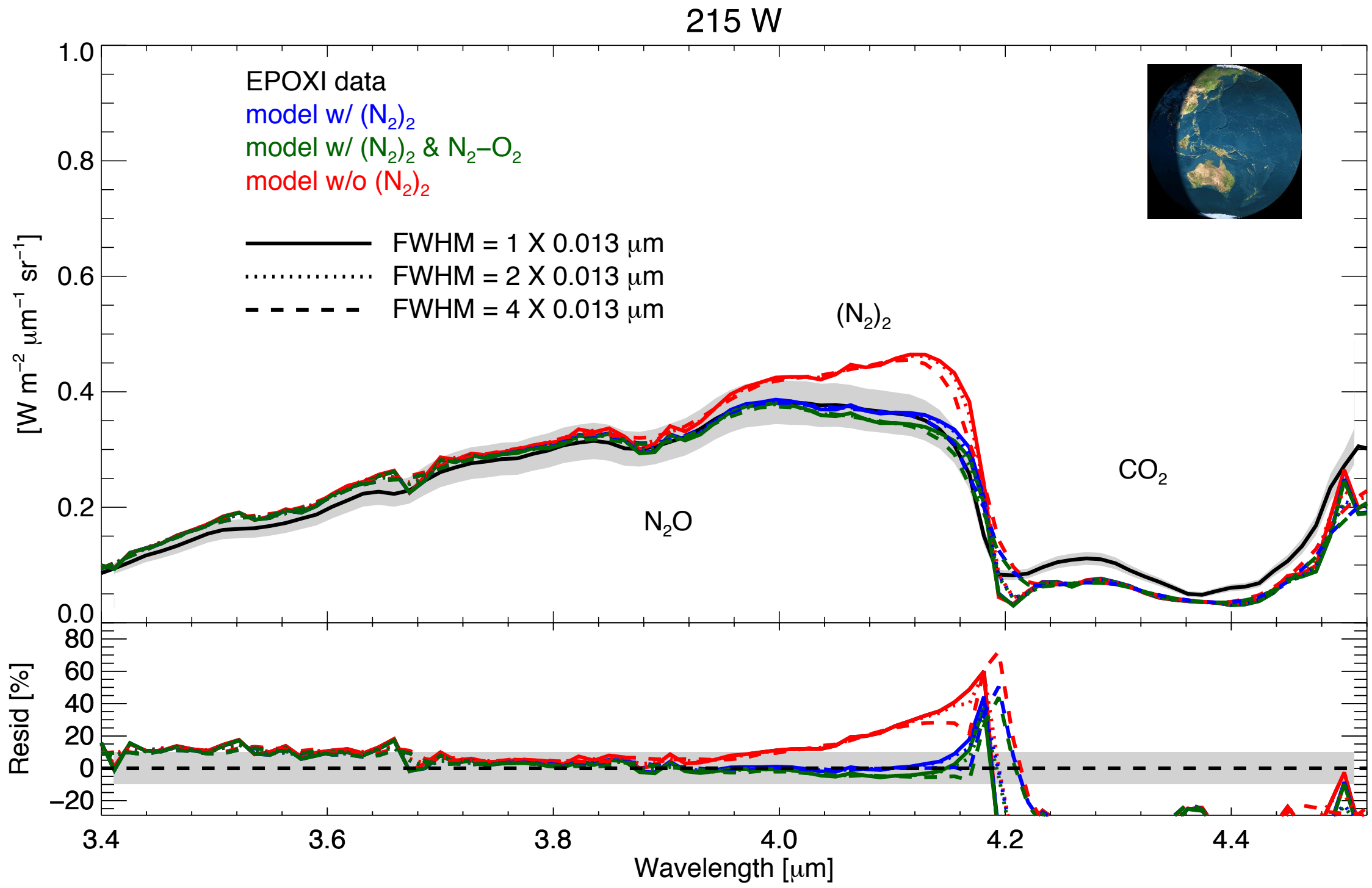

**Figure 4.** Spectral resolution sensitivity tests. The data are the same as in the upper left panel of Figure 2. Lines indicate Gaussian spectral convolution with different FWHMs in factors of 1 (solid line), 2 (dotted line), 4 (dashed line) × 0.013 um (the spectral bin size at $\lambda = 4.15$ μm reported by PDS).

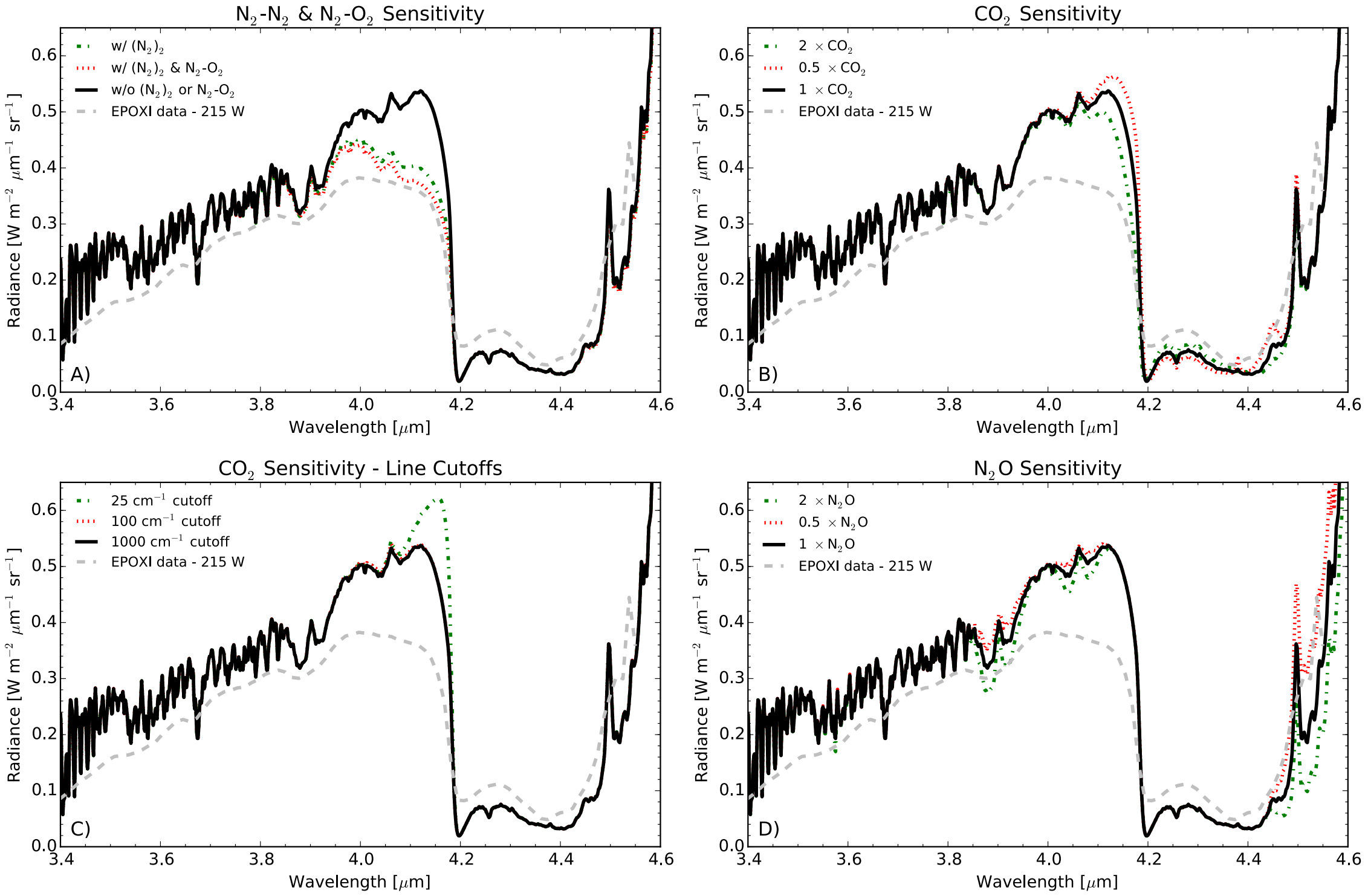

**Figure 5.** Sensitivity to molecular absorption by key species. Cloud-free spectra are generated by SMART with a solar zenith angle of 60°, an azimuth angle of 0°, and a Lambertian ocean surface. None of the synthetic spectra contain $N_2$-$N_2$ and $N_2$-$O_2$ absorption besides those in the top left panel. **A)** Model spectra without $(N_2)_2$ and $N_2$-$O_2$ absorption (black solid), with $(N_2)_2$ absorption only (green dash-dot), with $(N_2)_2$ and $N_2$-$O_2$ absorption (red dotted). **B)** Model spectra with 1 × standard $CO_2$ (black solid), 2 × standard $CO_2$ (green dash-dot), and 0.5 × standard $CO_2$ (red dotted). **C)** Model spectra with 1000 $cm^{-1}$ line cutoff for $CO_2$ absorption coefficients (black solid), with a 100 $cm^{-1}$ line cutoff (red dotted), and with a 25 $cm^{-1}$ line cutoff (green dash-dot). **D)** Model spectra with 1 × standard $N_2O$ (black solid), 2 × standard $N_2O$ (green dash-dot), and 0.5 × standard $N_2O$ (red dotted). Mixing ratio profiles assumed for $CO_2$ and $N_2O$ can be found in Figure 1. The *EPOXI* Earth spectrum with a sub-spacecraft longitude of 215° W (see Figure 3 and Table 1) is added to each panel for illustration (gray dashed), though it is not directly comparable with sensitivity spectra presented here.

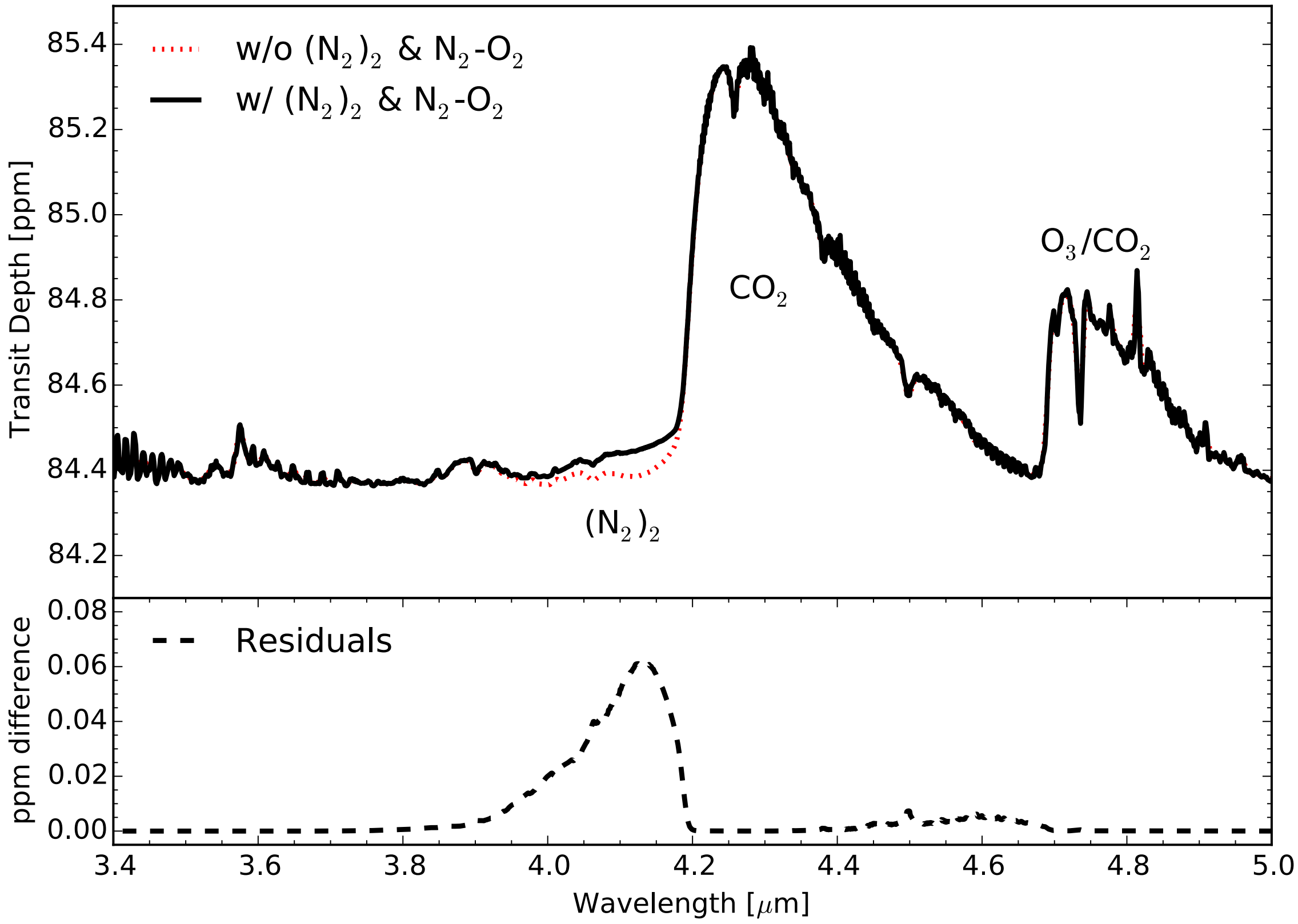

**Figure 6.** Transmission spectrum for an Earth transiting a Sun with (solid black) and without (dashed red) $(N_2)_2$ absorption in units of ppm. The spectral resolution is $\Delta\lambda$ = 0.005 μm. The bottom panel shows the difference between the model with and without $(N_2)_2$ absorption. The temperature and gas mixing ratio profiles assumed were the same as in Figure 1. Note the very small difference in transit depth is due to high limiting tangent altitudes for Earth-Sun geometries due to refraction, such that these spectra do not probe the deepest parts of the atmosphere where $(N_2)_2$ absorption is most significant (García Muñoz et al. 2012; Bétrémieux & Kaltenegger 2013; Misra et al. 2014b).

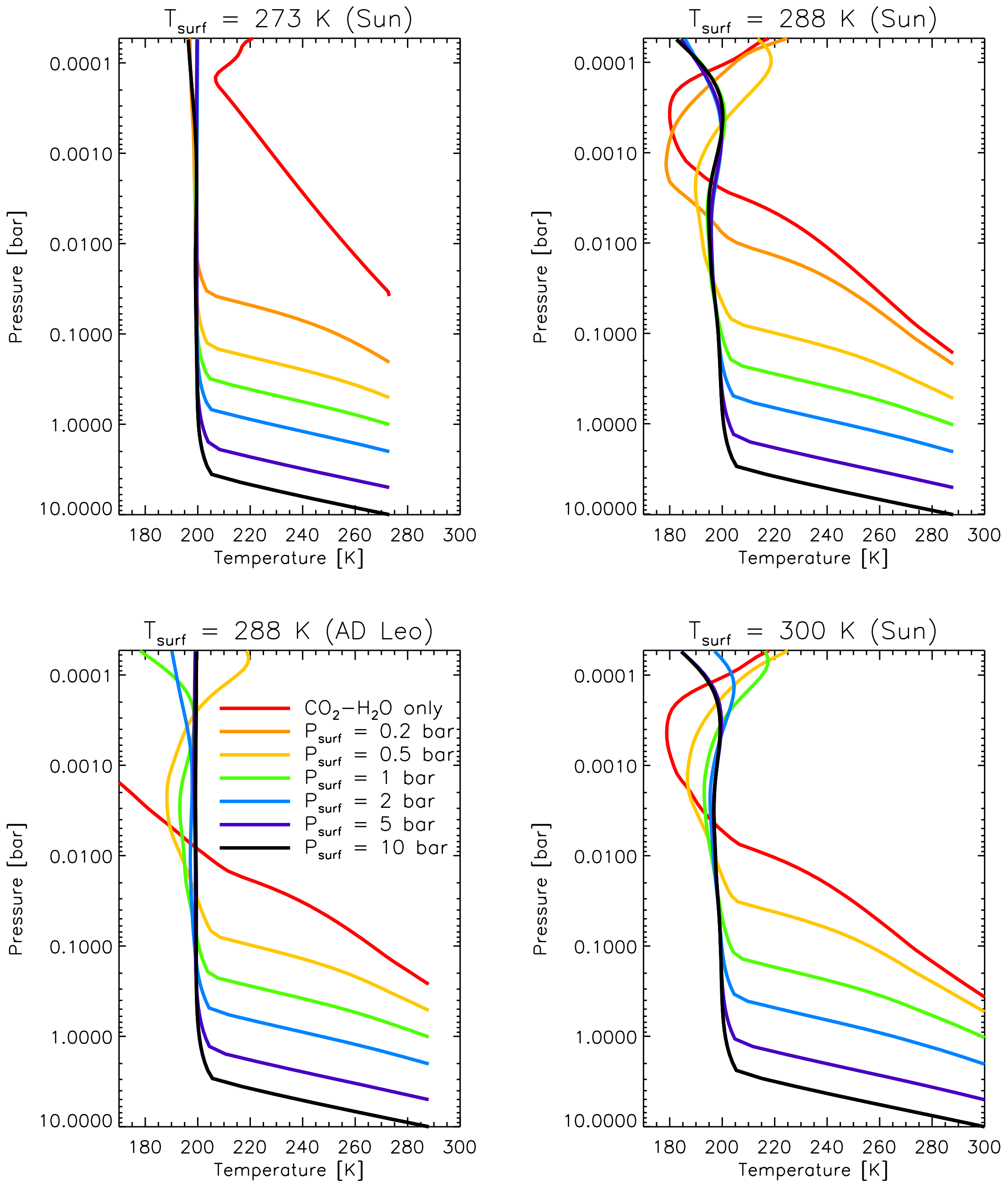

**Figure 7.** Self-consistent $N_2$-$CO_2$-$H_2O$ and $CO_2$-dominated atmosphere temperature-pressure profiles. The colors represent total surface pressures of 0.2 bar (orange), 0.5 bar (yellow), 1.0 bar (green), 2.0 bars (blue), 5 bars (purple), and 10 bars (black). The $CO_2$-dominated atmospheres that produce the same surface temperature are in red. More atmosphere parameters can be found in Table 2 and Table 3.

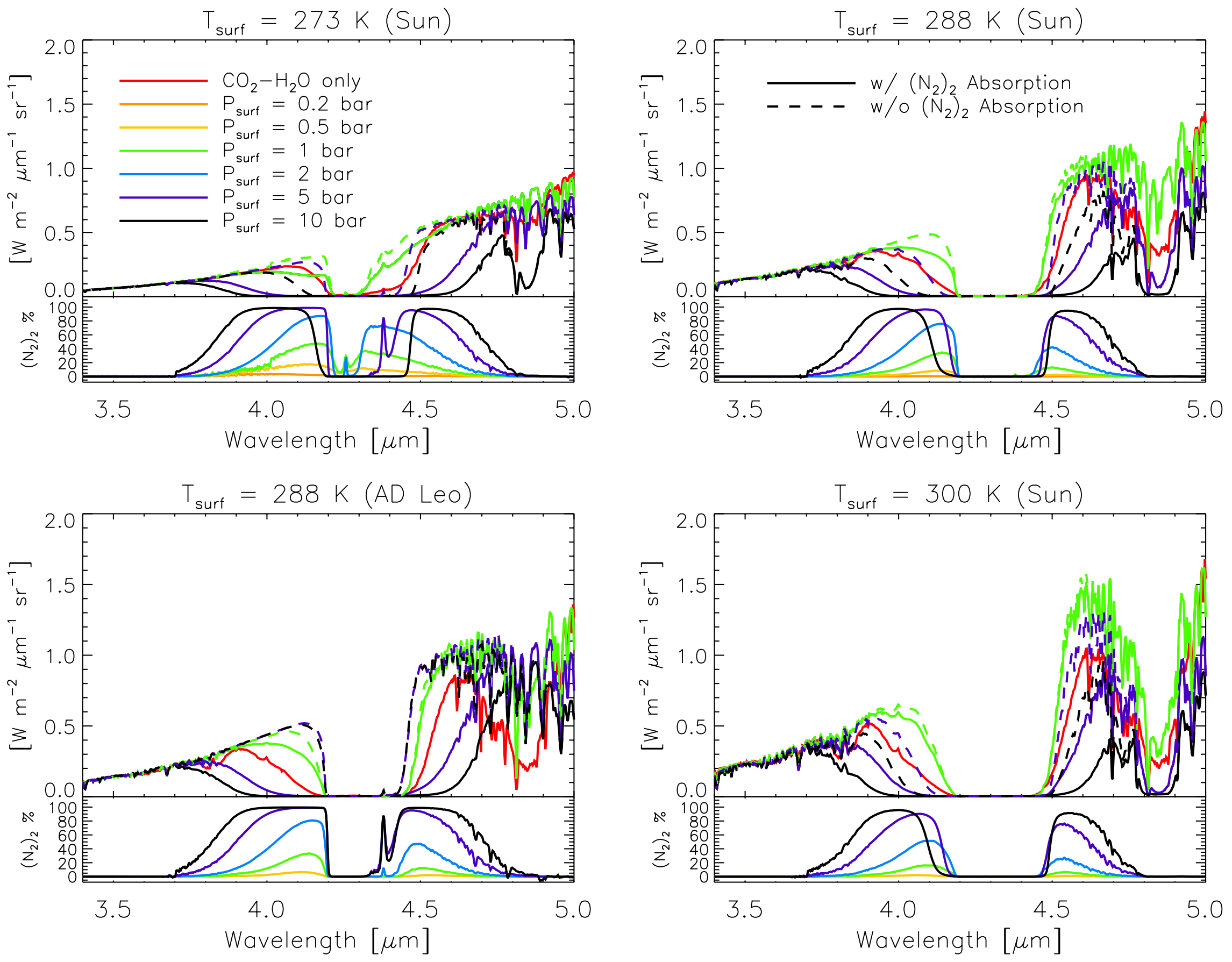

**Figure 8.** Thermal-only radiance spectra of model Earth-like planets showing $(N_2)_2$ absorption in the NIR. The temperature-pressure profiles used for these spectra are shown in Figure 7. In the top sub-panels only the $P_{surf}$ = 1, 5, and 10 bar, and the $CO_2$-dominated spectra are shown. The bottom sub-panels plot the wavelength-dependent percentage difference between the cases with $(N_2)_2$ absorption (solid) and without it (dashed) for all pressure scenarios given in the figure key.

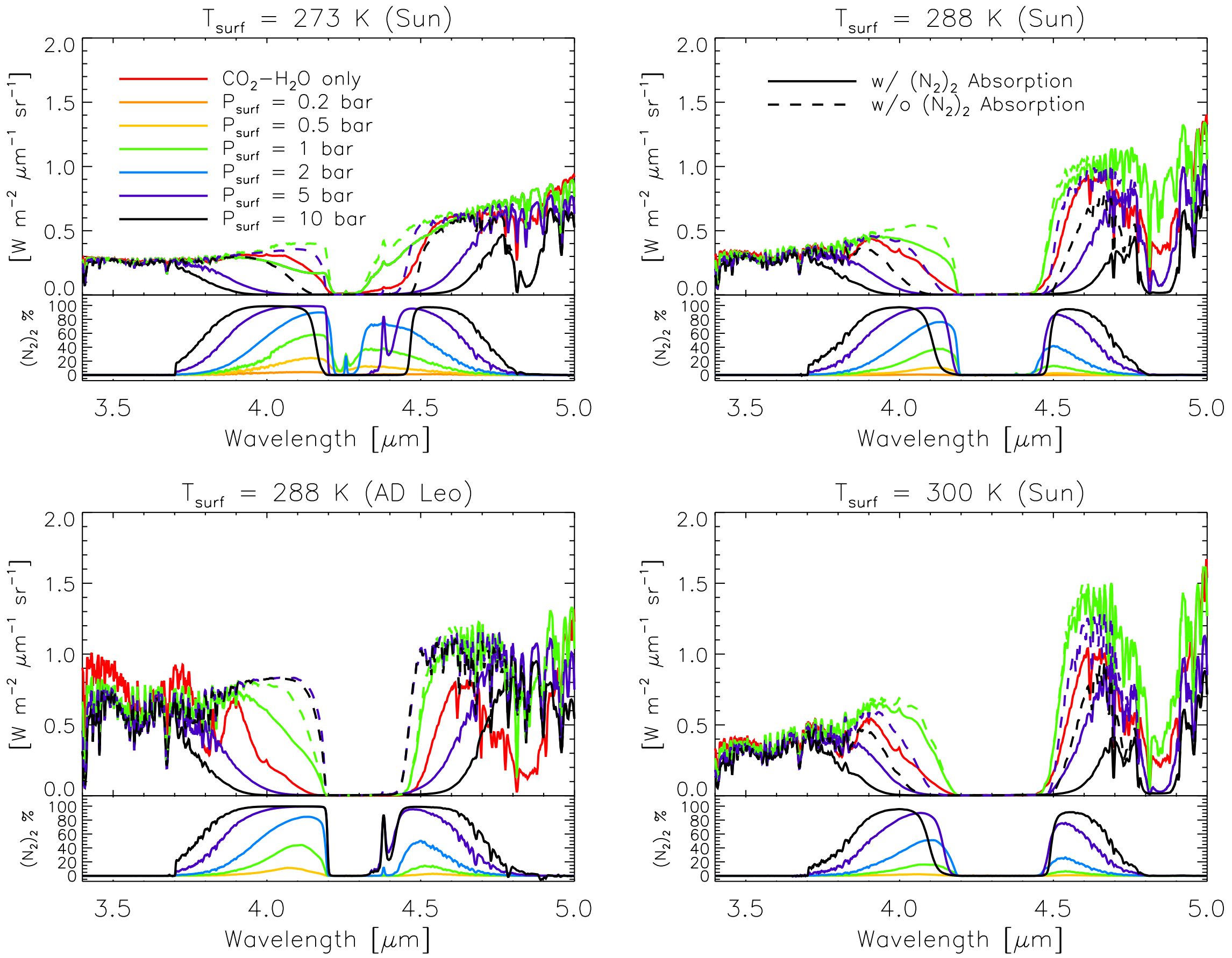

**Figure 9.** Same as Figure 8 except including both thermal emission and reflected light. The surface albedo is assumed to be 10% and constant over the wavelengths shown here. The incident stellar spectrum is either that of AD Leo (bottom left panel) or the Sun (all other panels).

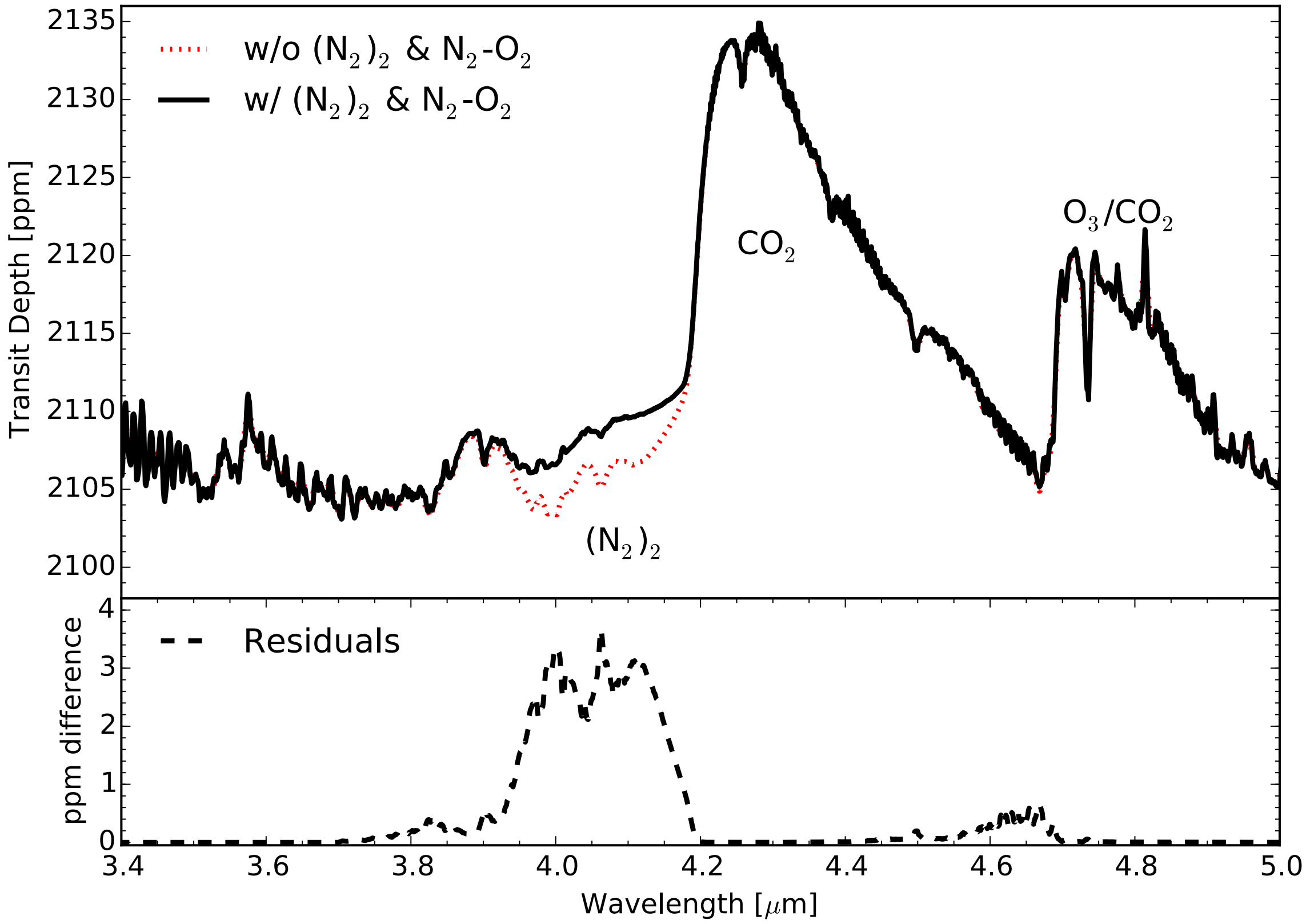

**Figure 10.** Transmission spectrum for an Earth transiting an M5V star (R = 0.2 R⊙; a = 0.05 AU) with (solid black) and without (dashed red) $(N_2)_2$ absorption in units of ppm. The spectral resolution is $\Delta\lambda = 0.005$ μm. The bottom panel shows the difference between the model with and without $(N_2)_2$ absorption. The temperature and gas mixing ratio profiles assumed were the same as in Figure 1.

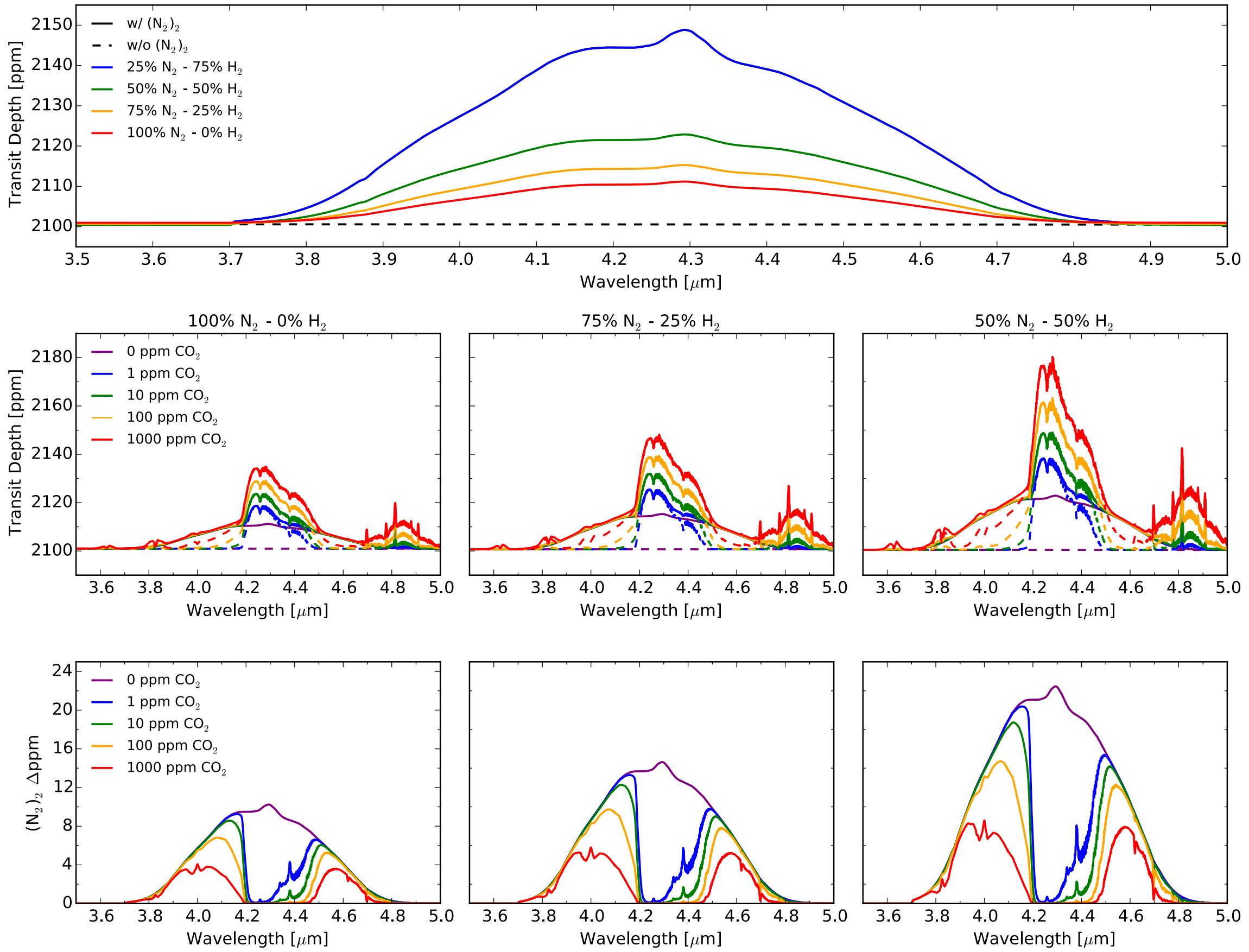

**Figure 11.** $(N_2)_2$ absorption in transmission for analytic $N_2$ and $N_2$-$H_2$-dominated atmospheres. Top panel: $(N_2)_2$ transmission depths for atmospheres that are pure $N_2$-$H_2$ mixtures (no $CO_2$). Middle panels: Transmission spectra for 100% $N_2$ (left), 75% $N_2$ (middle), and 50% $N_2$ (right) atmospheres with $CO_2$ concentrations of 0 ppm (purple), 1 ppm (blue), 10 ppm (green), 100 ppm (orange), and 1000 ppm (red) including $(N_2)_2$ absorption (solid lines) and without $(N_2)_2$ (dashed lines). Bottom panels: differences in transmission depths due to $(N_2)_2$ matched to spectra in the middle panel. The geometry assumed is a star with 20% the radius of the Sun, a planet-star distance of 0.05 AU and an impact parameter of 0.

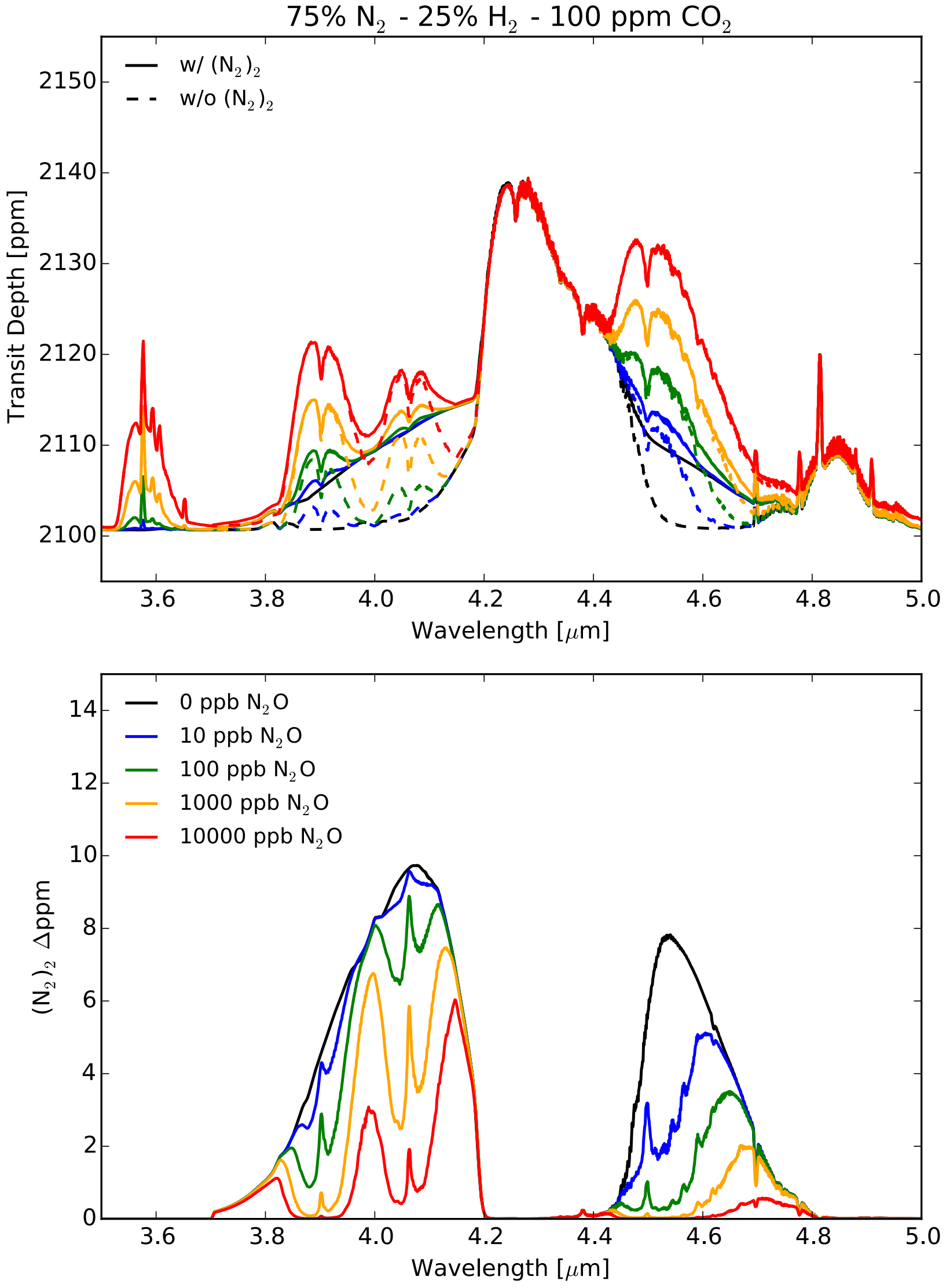

**Figure 12.** Sensitivity of $(N_2)_2$ absorption to different $N_2O$ abundances in transit transmission spectra. The atmosphere assumed is 75% $N_2$ and 25% $H_2$ with 100 ppm of $CO_2$ and 0 ppb (black), 10 ppb (blue), 100 ppb (green), 1000 ppb (orange), and 10000 ppb (red) of $N_2O$ (see Section 4). The spectral resolution is $\Delta\lambda$ = 0.005 μm. Top panels show the transmission spectra with $(N_2)_2$ (solid) or without $(N_2)_2$ (dashed) absorption. Bottom panels show differences in transmission depths due to $(N_2)_2$ matched to spectra in the top panel. The 0 ppb $N_2O$ case is identical to the 100 ppm $CO_2$ case in the lower middle

TABLES

**Table 1.** Summary of *EPOXI* Observations

| Midpoint time of observation (UT) | Sub-spacecraft latitude (°N) | Sub-spacecraft longitude (°W) | Sub-solar latitude (°S) | Sub-solar longitude (°W) |
|---|---|---|---|---|
| 2008-03-18 18:39:11 | 1.7 | 154.6 | 0.6 | 97.1 |
| 2008-03-18 22:39:11 | 1.6 | 214.6 | 0.5 | 157.1 |
| 2008-03-19 04:36:50 | 1.6 | 304.4 | 0.4 | 246.8 |
| 2008-03-19 12:36:50 | 1.5 | 64.6 | 0.3 | 6.8 |

**Table 2**. Parameters for $N_2$-$CO_2$-$H_2O$ atmospheres

| $P_{surf}$ [bar] | $f_{CO2}$ | Total $CO_2$ [bar] | Total $H_2O$ [bar] | Max $f_{H2O}$ | Planetary Albedo |
|---|---|---|---|---|---|
| $T_{surf}$ = 273 K (Sun) | | | | | |
| 0.2 | 7.90e-5 | 1.58e-5 | 2.32e-4 | 2.35e-2 | 0.223 |
| 0.5 | 1.90e-6 | 9.60e-7 | 4.41e-4 | 9.54e-3 | 0.238 |
| 1.0 | 1.10e-6 | 1.10e-6 | 7.43e-4 | 4.81e-3 | 0.260 |
| 2.0 | 1.05e-6 | 2.10e-6 | 1.31e-3 | 2.40e-3 | 0.291 |
| 5.0 | 6.50e-6 | 3.25e-6 | 3.00e-3 | 9.62e-4 | 0.353 |
| 10.0 | 1.10e-4 | 1.10e-3 | 5.82e-3 | 4.82e-4 | 0.412 |
| $T_{surf}$ = 288 K (Sun) | | | | | |
| 0.2 | 9.95e-2 | 1.99e-2 | 8.37e-4 | 6.25e-2 | 0.208 |
| 0.5 | 3.15e-3 | 1.58e-3 | 1.66e-3 | 2.62e-2 | 0.223 |
| 1.0 | 6.40e-4 | 6.40e-4 | 2.75e-3 | 1.33e-2 | 0.245 |
| 2.0 | 3.65e-4 | 7.30e-4 | 4.66e-3 | 6.72e-3 | 0.276 |
| 5.0 | 5.00e-4 | 2.50e-3 | 1.01e-2 | 2.70e-3 | 0.337 |
| 10.0 | 9.30e-4 | 9.30e-3 | 1.90e-2 | 1.35e-3 | 0.398 |
| $T_{surf}$ = 288 K (AD Leo)* | | | | | |
| 0.5 | 1.26e-2 | 6.30e-3 | 1.66e-3 | 2.62e-2 | 0.140 |
| 1.0 | 1.30e-3 | 1.30e-3 | 2.75e-3 | 1.33e-2 | 0.145 |
| 2.0 | 1.38e-4 | 2.76e-4 | 4.66e-3 | 6.71e-3 | 0.146 |
| 5.0 | 7.80e-6 | 3.90e-5 | 1.01e-2 | 2.70e-3 | 0.156 |
| 10.0 | 2.95e-6 | 2.95e-5 | 1.90e-2 | 1.35e-3 | 0.173 |
| $T_{surf}$ = 300 K (Sun)** | | | | | |
| 0.5 | 7.70e-2 | 3.85e-2 | 4.27e-4 | 5.32e-2 | 0.211 |
| 1.0 | 9.95e-3 | 9.95e-3 | 7.00e-3 | 2.75e-2 | 0.232 |
| 2.0 | 3.35e-3 | 6.70e-3 | 1.17e-2 | 1.40e-2 | 0.263 |
| 5.0 | 1.77e-3 | 8.85e-3 | 2.40e-2 | 5.65e-3 | 0.325 |
| 10.0 | 1.77e-3 | 1.77e-2 | 4.38e-2 | 2.85e-3 | 0.386 |

* For $P_{surf}$= 0.2 bar no $N_2$-$CO_2$-$H_2O$ atmosphere is possible; maximum $CO_2$ fraction produces $T_{surf}$ = 287.0K.

**For $P_{surf}$ = 0.2 bar no $N_2$-$CO_2$-$H_2O$ atmosphere is possible; maximum $CO_2$ fraction produces $T_{surf}$ = 284.5 K.

**Table 3**. Parameters for $CO_2$-dominated atmospheres

| $P_{surf}$ [bar] | $f_{CO2}$ | Total $CO_2$ [bar] | Total $H_2O$ [bar] | Max $f_{H2O}$ | Planetary Albedo |
|---|---|---|---|---|---|
| *$T_{surf}$ = 273 K (Sun)* | | | | | |
| 0.032 | 9.60e-1 | 3.07e-2 | 6.28e-5 | 1.38e-1 | 0.216 |
| *$T_{surf}$ = 288 K (Sun)* | | | | | |
| 0.148 | 9.60e-1 | 1.42e-1 | 6.66e-4 | 8.21e-2 | 0.213 |
| *$T_{surf}$ = 288 K (AD Leo)* | | | | | |
| 0.213 | 9.60e-1 | 2.04e-1 | 8.91e-4 | 5.90e-2 | 0.141 |
| *$T_{surf}$ = 300 K (Sun)* | | | | | |
| 0.332 | 9.60e-1 | 3.19e-1 | 3.07e-3 | 7.77e-2 | 0.213 |